\documentclass{jfm}
\usepackage{graphicx}
\usepackage{epstopdf, epsfig}

\usepackage{lmodern}
\usepackage[T1]{fontenc}
\usepackage[utf8]{inputenc}
\usepackage{indentfirst}
\usepackage{color}
\usepackage{microtype}
\usepackage{lipsum}	
\usepackage{parskip}
\usepackage{setspace}

\usepackage{graphicx}
\usepackage{subcaption}
\usepackage{epstopdf}
\usepackage{rotating}
\usepackage{morefloats}

\usepackage{longtable}
\usepackage{amsmath}
\usepackage{amssymb}
\usepackage{bm}
\usepackage{mnsymbol}
\usepackage{wasysym}
\usepackage{icomma}
\usepackage{multirow}
\usepackage{placeins}
\usepackage{mathrsfs}
\usepackage{mathtools}

\usepackage[english]{babel}
\usepackage{bibentry}
\usepackage{cleveref}
\usepackage{natbib}

\usepackage{makecell}

\shorttitle{Resolvent-based estimation of turbulent channel flow}
\shortauthor{F. R. Amaral, A. V. G. Cavalieri, E. Martini, P. Jordan and A. Towne}

\title{Resolvent-based estimation of turbulent channel flow using wall measurements}

\author{Filipe R. Amaral\aff{1}
  \corresp{\email{filipefra@ita.br}},
  André V. G. Cavalieri\aff{1},
	Eduardo Martini\aff{2},
	Peter~Jordan\aff{2}
	\and Aaron Towne\aff{3}}

\affiliation{\aff{1}Divisão de Engenharia Aeronáutica, Instituto Tecnológico de Aeronáutica, São José dos Campos, SP 12228-900, Brazil
							\aff{2}Département Fluides, Thermique, Combustion, Institut Pprime, CNRS–Université de Poitiers–ENSMA, 86000 Poitiers, France
							\aff{3}Department of Mechanical Engineering, University of Michigan, Ann Arbor, MI 48109, USA}

\begin{document}

\maketitle


\begin{abstract}

We employ a resolvent-based methodology to estimate velocity and pressure fluctuations within turbulent channel flows at friction Reynolds numbers of approximately 180, 550 and 1000 using measurements of shear stress and pressure at the walls, taken from direct numerical simulation (DNS) databases.
Martini \emph{et al.} (\emph{J. Fluid Mech.}, vol. 900, 2021, A2) showed that the resolvent-based estimator is optimal when the true space-time forcing statistics are utilized, thus providing an upper bound for the accuracy of any linear estimator.
We use this framework to determine the flow structures that can be linearly estimated from wall measurements, and we characterize these structures and the estimation errors in both physical and wavenumber space.
We also compare these results to those obtained using approximate forcing models -- an eddy-viscosity model and white-noise forcing -- and demonstrate the significant benefit of using true forcing statistics.
All models lead to accurate results up to the buffer-layer, but only using the true forcing statistics allows accurate estimation of large-scale log-layer structures, with significant correlation between the estimates and DNS results throughout the channel.
The eddy-viscosity model displays an intermediate behaviour, which may be related to its ability to partially capture the forcing colour.
Our results show that structures that leave a footprint on the channel walls can be accurately estimated using the linear resolvent-based methodology, and the presence of large-scale wall-attached structures enables accurate estimations through the logarithmic layer.

\end{abstract}


\begin{keywords}
Authors should not enter keywords on the manuscript, as these must be chosen by the author during the online submission process and will then be added during the typesetting process (see http://journals.cambridge.org/data/\linebreak[3]relatedlink/jfm-\linebreak[3]keywords.pdf for the full list)
\end{keywords}

\section{Introduction}
\label{sec:introduction}

The understanding of wall-bounded turbulent flow structures is a challenging research topic.
Flow structures of different natures and length scales are present in the flow and interact with one another.
\citet{jimenez2013near} presents a comprehensive review of near-wall turbulent structures, clarifying their role in momentum transfer and energy dissipation.
In the near-wall region, the main structures that arise are streamwise-velocity streaks; these structures contain most of the kinetic energy, and are accompanied by quasi-streamwise vortices, that organise the momentum transfer and energy dissipation.
Large-scale structures are present farther from the wall, where the flow here remains organised into streaks of larger size.
For high-Reynolds-number wall-bounded flows, very-large-scale motions (VLSM), or superstructures are found in the log layer \citep{smits2011high} and modulate the small-scale structures in the near-wall region \citep{marusic2010predictive, marusic2017scaling}.

The problem of flow state estimation using noisy, low-rank measurements has important research and industrial applications.
In experiments, one usually has only a limited number of probes available to characterise the flow field, which are often contaminated by measurement noise.
If a wall-bounded flow can be accurately estimated using only wall measurements, such as skin-friction and/or pressure, such estimates can be employed, for example, to feed a control scheme that would reduce drag and hence the required power in applications.
The review by \citet{bagheri2009input} addresses the problem of flow control employing an input-output formulation, which has several successful applications in delaying the transition to turbulence; see for instance \citet{morra2020realizable} and \citet{tol2019experimental} for recent implementations in simulation and experiment, respectively.
Moreover, the development of such prediction models and transfer functions between the available measurements and the flow state improve understanding of the physics of turbulent flows.

Estimation of channel flow using measurements of skin friction and pressure at the wall has been extensively investigated over the past two decades.
\citet{bewley2004skin} investigated a channel flow with direct numerical simulations (DNS) at friction Reynolds numbers of $Re_{\tau} = 100$ and 180.
Linear stochastic estimation (LSE) and adjoint-based algorithms were employed to estimate the flow field near the wall.
Good correlation between the actual flow field (obtained from DNS) and estimates were obtained for regions near the wall for $Re_{\tau} = 100$.
The correlation between the DNS and estimated streamwise velocity was greater than 0.9 up to 20 wall units into the channel ($y^+ \leq 20$).
On the other hand, in the center of the domain, the correlation approached 0 for the LSE algorithm and 0.1 for the adjoint-based methodology.
\citet{hoepffner2005state} and \citet{chevalier2006state} performed low Reynolds number ($Re_{\tau} = 100$) channel flow state estimation based on the linearised Navier-Stokes (LNS) operator for both laminar and turbulent flows, respectively.
The channel flow databases were obtained through DNS and estimates were obtained from noisy measurements of wall pressure and skin friction using a Kalman filter.
Good agreement between DNS and estimates was observed. 
An important feature in \citet{chevalier2006state} was the inclusion of spatial statistics of the non-linear terms in the LNS system, although those statistics were considered white in time.
The use of spatial statistics improved the accuracy of the estimates.
Later, \citet{martini2020resolvent} showed that using spatio-temporal statistics provides considerable improvements on top of those obtained by \citet{chevalier2006state}.

\citet{colburn2011state} extended these earlier works by using an ensemble Kalman filter to estimate a turbulent channel flow.
While the Kalman filter is based on the LNS operator, the ensemble Kalman filter allows consideration of the non-linear dynamics.
The estimation quality by \citet{colburn2011state} was improved by at least one order of magnitude in the near-wall region.
Near-perfect estimates were obtained very close to the wall ($y^{+} \leq 20$).

Moving to higher Reynolds numbers, \citet{illingworth2018estimating} used a Kalman filter based on the LNS equations forced by white noise to build a linear estimator for a DNS of channel flow at $Re_{\tau} = 1000$.
Instead of wall quantities, the authors observed time-resolved velocities at a single wall-normal distance and attempted to estimate the velocity field at other wall-normal distances.
In comparison with DNS results, the amplitude of the velocity fluctuations were overpredicted, although the large scale structures were, qualitatively, well estimated.
The results were improved by inclusion of an eddy-viscosity model \citep{pujals2009note, reynolds1967stability}, leading to better agreement between estimate and DNS fluctuations.
Their approach, including the use of an eddy viscosity model, was applied to a $Re_{\tau} = 2000$ database in \citet{oehler2018linear}.
The authors conducted estimates using the velocity components at a single wall-normal distance as input, as in \citet{illingworth2018estimating}, and also using the wall shear-stress.
Similar performance as that obtained for $Re_{\tau} = 1000$ was achieved when using the velocity components measured at a distance $y/H = 0.2$, where $y$ is the the wall-normal coordinate and $H$ is the channel half-height.
On the other hand, the estimator performance deteriorated when the wall skin friction was used as input.

The aforementioned works dealt mainly with model-based estimation.
However, it is also possible to perform flow estimations based solely on data. 
\citet{encinar2019logarithmic} used LSE and noiseless wall measurements to estimate the velocity components of turbulent channel flow for friction Reynolds numbers ranging from 932 to 5300.
The velocity components were well estimated in the buffer layer, but only large-scale structures were captured farther from the wall.
Neural networks trained using DNS data were employed by \citet{guastoni2020convolutional} to estimate an open channel flow based on wall quantities at $Re_{\tau} = 180$ and 550.
The authors used two different algorithms: a fully-convolutional neural network (FCN) and a fully-convolutional neural network based on proper orthogonal decomposition (FCN-POD).
The first algorithm (FCN) performed better in estimating the flow near the wall, whereas the second (FCN-POD) outperformed FCN farther from the wall.
The $Re_{\tau} = 180$ database was also employed to train the FCN model and estimate the $Re_{\tau} = 550$ flow field and good agreement between DNS and estimates was obtained up to $y^{+} = 50$.
In a study of turbulent boundary-layers, \citet{sasaki2019transfer} used velocity fluctuation measurements at different wall-normal positions in order to estimate the streamwise velocity component of the flow field, building from a technique developed for turbulent jets \citep{sasaki2017real}.
The authors compared the ability of three methods to estimate the flow field: a single input linear transfer function, a multiple-input linear transfer function, and a non-linear transfer function.
Linear transfer functions were obtained using data from large-eddy simulations, in a process equivalent to LSE in the frequency domain \citep{tinney2006spectral}.
Off-design predictions for different Reynolds numbers were also addressed.
Results showed good agreement between simulation data and estimates.

Compared to the aforementioned data-based methods, model-based estimation methods have the advantage of providing, in addition to the estimated fields, a theoretical framework to understand the properties of estimated structures.
Linearised models for turbulent flows can be obtained using the resolvent framework.
Resolvent analysis employs the LNS equations and models the non-linear terms as a stochastic forcing term \citep{mckeon2010critical, hwang2010linear, beneddine2016conditions, taira2017modal}.
In this framework, the non-linear (forcing) terms in the Navier-Stokes system generate a linear response through the resolvent operator, providing an input-output formulation.
The input thus corresponds to the forcing terms and the output denotes the flow state.
Moreover, it is possible to obtain modes by a singular-value decomposition of the resolvent operator, the so-called resolvent modes, that provide optimal forcing and response modes.
\citet{towne2018spectral} demonstrated a connection between the spectral proper orthogonal decomposition (SPOD) and resolvent response modes; the response modes are equivalent to the SPOD modes when analyzing the flow response to stochastic white-noise forcing.
Resolvent analysis has been extensively employed to characterise wall-bounded and free shear flows \citep{mckeon2010critical, hwang2010linear, sharma2013coherent, abreu2020spod, abreu2020resolvent, morra2021colour, jeun2016input, abreu2017coherent, towne2017statistical, schmidt2018spectral, lesshafft2019resolvent}.

A resolvent-based estimation framework was introduced by \citet{towne2020resolvent}, where the main objective was to estimate space-time flow statistics from limited measurements.
The central idea is to use the measurements to approximate the non-linear terms that act as a forcing on the LNS equations, which in turn provide an estimate of the flow state upon application of the resolvent operator in the frequency domain.
\citet{martini2020resolvent} extended this resolvent-based methodology to obtain optimal, non-causal estimates of time-varying flow quantities. 
Just as in the derivation of the Kalman filter, the optimal resolvent-based transfer functions between the measurements and the state depend on \emph{a priori} knowledge of the forcing statistics.  
However, the resolvent-based methodology can account for forcing with space-time colour, which cannot be easily handled by a Kalman filter.
Estimation of a DNS of channel flow at $Re_{\tau} = 180$, using wall measurements, provided the velocity components for a wavenumber-frequency combination corresponding to the near-wall cycle of streamwise vortices and streaks.
Overshoots in the streamwise velocity were observed when insufficient information on forcing statistic was used, as previously reported by \citet{chevalier2006state} and \citet{illingworth2018estimating}, but close agreement between DNS and simulation results was achieved when the true forcing colour (the complete two-point space-time statistics) was used.

In the present study, the resolvent-based methodology first introduced by \citet{martini2020resolvent} is employed to estimate the space-time field of turbulent channel flow from low-rank wall measurements.
This is pursued for turbulent channel flow with $Re_{\tau} \approx 180$, 550 and 1000, and full fields are estimated with the resolvent-based approach for the first time.
By using DNS databases, this work allows an assessment of the accuracy of resolvent-based estimation for turbulent flows with increasing Reynolds number, showing the range of scales that can be estimated from wall measurements.
It is clear from the above discussion that the use of linearised models allows modelling several dynamical features of turbulent flows, and research remains active on this subject.
For instance, different choices of linearised operator (including or not an eddy-viscosity) have been explored by \citet{morra2019relevance} and \citet{symon2021energy}, and other possible linearisations were explored by \citet{farrell2012dynamics} and \citet{thomas2015minimal}, with a streamwise-averaged mean flow, changing in span and in time, being used to build a system with restricted non-linearity, reproducing many features of turbulent flows.
The present work follows a different approach to obtain flow estimations from wall quantities.
Instead of studying how different linearisations affect estimation performance, we investigate the impact of including accurate space-time statistics of the non-linear terms in the construction of the estimator.
The resolvent-based approach of \citet{martini2020resolvent} uses these statistics to produce the optimal linear estimator.
In this paper, we investigate how much of the flow can be estimated, in both physical and wavenumber space, using this optimal linear estimator and measurements at the wall.
For completeness, the work also presents simpler estimations considering white-noise forcing of the linearised operators, with or without an eddy-viscosity model.
We show that including accurate space-time statistics of the nonlinear terms, made possible by the resolvent-based estimation framework, significantly improves the estimation accuracy compared to the typical white-noise assumption.

The remainder of the manuscript is organised as follows.
The methods are introduced in \S \ref{sec:methodology}, which includes a brief description the of transfer function used to estimate the flow state.
The channel flow DNS details are also addressed in \S \ref{sec:methodology}.
In \S \ref{sec:Retau550} we apply the resolvent-based methodology to estimate the flow state and forcing components of the $Re_{\tau} \approx 550$ flow.
Detailed results are provided, focusing on the different forcing statistics models and choice of sensors.
Three strategies for modelling the nonlinear forcing terms are considered: computing their true values from DNS data, assuming white noise, and including an eddy viscosity model.
Different choices of sensors are explored through comparisons among estimates obtained using skin friction and/or wall pressure on one of both sides of the channel.
Qualitative snapshots of state and forcing components from the DNS and estimates are shown and quantitative metrics are introduced in order to evaluate the quality of the estimates and the power spectra of flow fluctuations.
In the sequence, \S \ref{sec:Retau180_1000} addresses the effect of Reynolds number, including estimator performance comparisons with literature \citet{illingworth2018estimating, oehler2018linear}.
The three Reynolds numbers are studied to determine the range of scales that may be accurately estimated using wall measurements.
Finally, \S \ref{sec:conclusions} summarizes the conclusions.

\section{Methodology}
\label{sec:methodology}

\subsection{Resolvent-based estimation}
\label{sec:resolvent}

Consider the linearised Navier-Stokes (LNS) equations and the flow state vector $\boldsymbol{q} = [\boldsymbol{u}~\boldsymbol{v}~\boldsymbol{w}~\boldsymbol{p}]^T$, where $\boldsymbol{u}$, $\boldsymbol{v}$ and $\boldsymbol{w}$ are the streamwise, wall-normal and spanwise velocity components, $\boldsymbol{p}$ is the pressure component, and $T$ denotes transpose.
The flow state is decomposed according to the Reynolds decomposition $\boldsymbol{q} = \boldsymbol{\bar{q}} + \boldsymbol{q^{\prime}}$, were $\boldsymbol{\bar{q}}$ is the mean flow (or, more generally, any base flow) and $\boldsymbol{q^{\prime}}$ is the fluctuation around the base flow.
The forcing $\boldsymbol{f}$ is obtained by gathering terms in the Navier-Stokes system that are non-linear in the fluctuation $\boldsymbol{q}$.
This decomposition is based on the LNS system considering only a molecular viscosity, and is exact if the forcing $\boldsymbol{f}$ is calculated considering all non-linear terms.
In the following, primes ($^{\prime}$) will be omitted for notational simplicity.
For the channel flow configuration employed in this study, the linearised equations are written separately for each streamwise and spanwise wavenumber, such that the discretised state has a dimension equal to $4 N_y$, where $N_y$ is the number of wall-normal points.
Taking advantage of the periodicity, Fourier transforms are applied in the streamwise and spanwise directions, i.e.,
\begin{subeqnarray}
	\boldsymbol{q}(\alpha,y,\beta,t) &=& \int^{\infty}_{-\infty} \int^{\infty}_{-\infty} \boldsymbol{q}(x,y,z,t) e^{-i \alpha x -i \beta z} dx~dz \mbox{,} \\
	\boldsymbol{f}(\alpha,y,\beta,t) &=& \int^{\infty}_{-\infty} \int^{\infty}_{-\infty} \boldsymbol{f}(x,y,z,t) e^{-i \alpha x -i \beta z} dx~dz \mbox{,}
	\label{eq:fft_xz}
\end{subeqnarray}
\noindent where $\alpha$ and $\beta$ denote the streamwise and spanwise wavenumbers, respectively, $x$, $y$ and $z$ indicate the streamwise, wall-normal and spanwise directions, respectively, and $t$ is time.

Writing the LNS equations in a discretised state-space form, considering a grid with $N_y$ points in the wall-normal direction, one has
\begin{subeqnarray}
	\mathsfbi{M} \frac{d \boldsymbol{q}(t)}{d t} &=& \mathsfbi{A} \boldsymbol{q}(t) + \mathsfbi{B} \boldsymbol{f} \mbox{,} \\
	\boldsymbol{y}(t) &=& \mathsfbi{C} \boldsymbol{q}(t) + \boldsymbol{n}(t) \mbox{,}
	\label{eq:state-space_time}
\end{subeqnarray}
\noindent where $\mathsfbi{A} \in \mathbb{C}^{4 N_{y} \times 4 N_{y}}$ is the linear operator, $\mathsfbi{B} \in \mathbb{C}^{4 N_{y} \times 3 N_{y}}$ is the input matrix that restricts the forcing terms to appear only in the momentum equation, $\boldsymbol{y} \in \mathbb{C}^{N_{s}}$ is the system observation, $\mathsfbi{C} \in \mathbb{C}^{N_{s} \times 4 N_{y}}$ is the observation matrix that selects $N_{s}$ sensor readings from the state vector, and $\boldsymbol{n} \in \mathbb{C}^{N_{s}}$ is the measurement noise.
$\mathsfbi{M} \in \mathbb{C}^{4 N_{y} \times 4 N_{y}}$ is a diagonal matrix whose entries are set to one and zero for the momentum equations and continuity equations, respectively.
Dependency on wavenumbers $\alpha$ and $\beta$, as well as on wall-normal variable $y$, was dropped to simplify notations.
The flow forcing variables comprise $\boldsymbol{f} = [\boldsymbol{f_x}~\boldsymbol{f_y}~\boldsymbol{f_z}]^T \in \mathbb{C}^{3 N_{y}}$, where $\boldsymbol{f_x}$, $\boldsymbol{f_y}$ and $\boldsymbol{f_z}$ are the streamwise, wall-normal and spanwise forcing components, respectively (see App. \ref{app:math} for details).
If the forcing $\boldsymbol{f}$ in (\ref{eq:state-space_time}a) is defined as the non-linear perturbation terms in the Navier-Stokes system, the relation in (\ref{eq:state-space_time}a) is exact \citep{mckeon2010critical}.

In the frequency domain, (\ref{eq:state-space_time}) is written as
\begin{equation}
	\boldsymbol{\hat{y}}(\omega) = \left[\mathsfbi{C} (- i \omega \mathsfbi{M} - \mathsfbi{A})^{-1} \mathsfbi{B}\right] \boldsymbol{\hat{f}}(\omega) + \boldsymbol{\hat{n}}(\omega) \mbox{,}
	\label{eq:state-space_freq}
\end{equation}
\noindent where 
\begin{equation}
	\boldsymbol{\hat{y}} = \int \limits_{-\infty}^{\infty} \boldsymbol{y}(t) e^{i \omega t} dt
	\label{eq:fourier}
\end{equation}
is the Fourier transform of $\boldsymbol{y}$, with analogous expression for other variables.
$\omega$ is the temporal frequency and $i = \sqrt{-1}$.
The term $\mathsfbi{R} = (- i \omega \mathsfbi{M} - \mathsfbi{A})^{-1}$ is the resolvent operator, i.e., the transfer function between the forcing terms and the flow response \citep{mckeon2010critical, cavalieri2019wave, towne2020resolvent}.
Appendix \ref{app:math} contains the expressions for the linear operator $\mathsfbi{L} = (- i \omega \mathsfbi{M} - \mathsfbi{A})$, input matrix $\mathsfbi{B}$ and observation matrix $\mathsfbi{C}$.

In the present study, we consider the observations to wall shear-stress components ($\frac{du}{dy}\rvert_{wall}$ and $\frac{dw}{dy}\rvert_{wall}$) and pressure ($p_{wall}$) on one or both walls for $Re_{\tau} \approx 180$ and $\approx 550$ cases (6 sensors for each wavenumber), whereas for the $Re_{\tau} \approx 1000$ case, only wall shear stresses are available (4 sensors for each wavenumber).
As data is obtained directly from the DNS, measurement noise is non-existent.
However, as will be shown latter, measurement noise regularises the estimation, and thus the measurement noise $\boldsymbol{\hat{n}}$ was taken as a small value close to machine precision.
This is analogous to the ill-posedness of Kalman-filter estimation for vanishing sensor noise.   
The small values are used in order to approach the zero-noise limit, justified by the high fidelity of the database, while keeping the estimation problem well posed.

Following \citet{martini2020resolvent}, the optimal linear transfer function $\boldsymbol{\hat{T}_f}$ between the system observation $\boldsymbol{\hat{y}}$ and the estimated flow forcing $\boldsymbol{\hat{\tilde{f}}}$,
\begin{equation}
	\boldsymbol{\hat{\tilde{f}}} = \boldsymbol{\hat{T}_f} \boldsymbol{\hat{y}} \mbox{,}
	\label{eq:f_est}
\end{equation}
\noindent obtained by minimizing the error between the true ($\boldsymbol{\hat{f}}$) and estimated ($\boldsymbol{\hat{\tilde{f}}}$) forcings, takes the form
\begin{equation}
	\boldsymbol{\hat{T}_f} = \mathsfbi{P_{ff}} {\mathsfbi{R_y}}^{*} \left(\mathsfbi{R_y} \mathsfbi{P_{ff}} {\mathsfbi{R_y}}^{*} + \mathsfbi{P_{nn}}\right)^{-1} \mbox{.}
	\label{eq:Tf}
\end{equation}

In (\ref{eq:Tf}), $\mathsfbi{P_{ff}} = \langle \boldsymbol{\hat{f}} \boldsymbol{\hat{f}}^* \rangle$ and $\mathsfbi{P_{nn}} = \langle \boldsymbol{\hat{n}} \boldsymbol{\hat{n}}^* \rangle$ are the cross spectral densities (CSDs) of the flow forcing and measurement noise statistics, respectively, and $\mathsfbi{R_y} = \mathsfbi{C} \mathsfbi{R} \mathsfbi{B}$ is the resolvent operator relating the forcing terms to the sensor readings.
Here, $\langle \cdot \rangle$ denotes an ensemble average and the asterisk indicates a Hermitian transpose.
If $\mathsfbi{P_{ff}}$ is not known, which is frequently the case, it must be replaced by an \textit{a priori} ansatz in order to evaluate the transfer function $\boldsymbol{\hat{T}_f}$.
Note that the forcing statistics are used to estimate instantaneous forces from sensors.
An optimal linear estimator is obtained only if $\mathsfbi{P_{ff}}$ is known \citep{martini2020resolvent}.
This, however, is typically not the case.
For our estimates, we will consider three ans\"{a}tze for the forcing CSD: the true CSD computed from DNS data, $\mathsfbi{P_{ff,DNS}}$; a spatially temporal white-noise assumption, $\mathsfbi{I}$, and an eddy viscosity model to (partially) account for the unknown forces (see App. \ref{app:math} for the eddy-viscosity model linear operator formulation).
Alternatively, \citet{martini2020resolvent} suggested a method to model the forcing statistics using additional sensors, but this is not pursued here.

Given the resolvent input-output formulation, the optimal state estimation is obtained from the optimal force estimation, i.e.,
\begin{equation}
	\boldsymbol{\hat{\tilde{q}}} = \mathsfbi{R} \mathsfbi{B} \boldsymbol{\hat{\tilde{f}}} \mbox{,}
	\label{eq:q_est}
\end{equation}
\noindent where $\boldsymbol{\hat{\tilde{q}}}$ is the estimated flow state.
The transfer function relating the system observation $\boldsymbol{\hat{y}}$ and the flow estimated state $\boldsymbol{\hat{\tilde{q}}}$ is thus obtained as
\begin{equation}
	\boldsymbol{\hat{T}_q} = \mathsfbi{R} \mathsfbi{B} \boldsymbol{\hat{T}_f} \mbox{,}
	\label{eq:Tq}
\end{equation}
\noindent hence,
\begin{equation}
	\boldsymbol{\hat{\tilde{q}}} = \boldsymbol{\hat{T}_q} \boldsymbol{\hat{y}} \mbox{.}
	\label{eq:q_est2}
\end{equation}

It is worth mentioning that the expression obtained for the transfer function $\boldsymbol{\hat{T}_f}$ is closely related to Wiener filter estimation \citep{meditch1973survey, martinelli2009feedback}.
More details and further discussion of the method employed to derive the estimator, including strategies for efficient computational implementation for large systems, can be found in \citet{martini2020resolvent}.

Taking inverse Fourier Transforms of (\ref{eq:Tf}) and (\ref{eq:Tq}), in order to return to time domain, leads to
\begin{subeqnarray}
	\boldsymbol{T_f}(\alpha,y,\beta,t) &=& \int^{\infty}_{-\infty} \boldsymbol{\hat{T}_f}(\alpha,y,\beta,\omega) e^{i \omega t} d\omega \mbox{,} \\
	\boldsymbol{T_q}(\alpha,y,\beta,t) &=& \int^{\infty}_{-\infty} \boldsymbol{\hat{T}_q}(\alpha,y,\beta,\omega) e^{i \omega t} d\omega \mbox{.}
	\label{eq:ifft_TfTq_time}
\end{subeqnarray}

It is then possible to reconstruct the snapshots of the forcing and state estimates by performing a convolution of the transfer functions $\boldsymbol{T_f}$ and $\boldsymbol{T_q}$ with the measurements $\boldsymbol{y}$,
\begin{subeqnarray}
	\boldsymbol{\tilde{f}}(\alpha,y,\beta,t) &=& \int^{\infty}_{-\infty} \boldsymbol{T_f}(\alpha,y,\beta,\tau) \boldsymbol{y}(\alpha,y,\beta,t-\tau) d\tau \mbox{,} \\
	\boldsymbol{\tilde{q}}(\alpha,y,\beta,t) &=& \int^{\infty}_{-\infty} \boldsymbol{T_q}(\alpha,y,\beta,\tau) \boldsymbol{y}(\alpha,y,\beta,t-\tau) d\tau \mbox{,}
	\label{eq:fq_time}
\end{subeqnarray}
\noindent respectively.

Inverse Fourier Transforms were also performed in the streamwise and spanwise directions to the estimates in space and time, i.e.,
\begin{subeqnarray}
	\boldsymbol{\tilde{f}}(x,y,z,t) &=& \int^{\infty}_{-\infty} \int^{\infty}_{-\infty} \boldsymbol{\tilde{f}}(\alpha,y,\beta,t) e^{i \alpha x + i \beta z} d\alpha~d\beta \mbox{,} \\
	\boldsymbol{\tilde{q}}(x,y,z,t) &=& \int^{\infty}_{-\infty} \int^{\infty}_{-\infty} \boldsymbol{\tilde{q}}(\alpha,y,\beta,t) e^{i \alpha x + i \beta z} d\alpha~d\beta \mbox{.}
	\label{eq:ifft_ab}
\end{subeqnarray}

The formulation above may also be entirely written in space and time, with forcing and state estimated by a space-time convolution between $\boldsymbol{\hat{y}}$ and the transfer functions.
However, the use of periodic boundary conditions in the channel-flow simulations allow the evaluation of each wavenumber separately, which is more computationally efficient. 

As a final remark, note that the convolutions in (\ref{eq:fq_time}) lead to non-causal estimations.
This is intrinsic to the frequency-domain approach, and prevents its direct application in real time, as is necessary for flow control.
The goal of this work to thoroughly investigate which flow structures may be accurately estimated with a linear model from wall sensors.
Thus, without restrictions related to causality, we here evaluate which turbulent structures leave a wall imprint allowing them to be estimated from a time series of wall measurements.
A causal extension is nonetheless possible via the Wiener-Hopf formalism, as discussed in \cite{martinelli2009feedback} and \cite{martini2019linear}.

\subsection{Numerical simulations}
\label{sec:dns}

\begin{figure}
	\centerline{\includegraphics[width=0.6\textwidth]{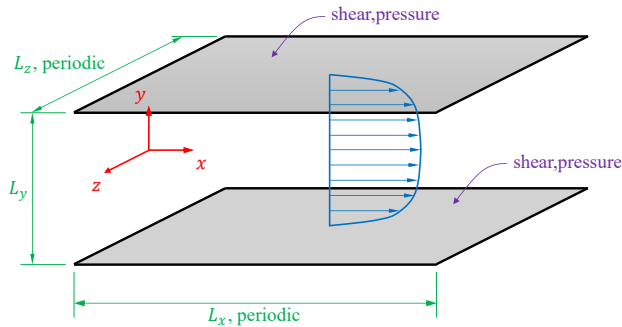}}
	\caption{Sketch of the channel flow geometry, coordinate system (red), mean flow (blue) and wall measurements employed to perform the estimations (gray).}
	\label{fig:channel_flow_sketch}
\end{figure}

The DNS were conducted with the ChannelFlow pseudo-spectral code \citep{gibson2019channelflow}.
Figure \ref{fig:channel_flow_sketch} shows a sketch of the geometry and coordinate system of the problem.
The box dimensions were $2\pi \times 2 \times \pi$ in the streamwise ($L_x$), wall-normal ($L_y$) and spanwise ($L_z$) directions, respectively, for the $Re_{\tau} \approx 550$ and 1000 cases, whereas a box of $4\pi \times 2 \times 2\pi$ dimensions was employed for the $Re_{\tau} \approx 180$ simulation.
Periodic boundary conditions are assumed in the streamwise and spanwise directions.
The wall-normal direction ($y$) was discretised with Chebyshev polynomials, using the same number of polynomials as other simulations \citep{lozanoduran2014effect}, and the streamwise ($x$) and spanwise ($z$) directions were discretised using Fourier modes, including de-aliasing \citep{delalamo2004scaling}.

\begin{figure}
	\centering
	\begin{subfigure}{0.48\textwidth}
		\includegraphics[width=\textwidth]{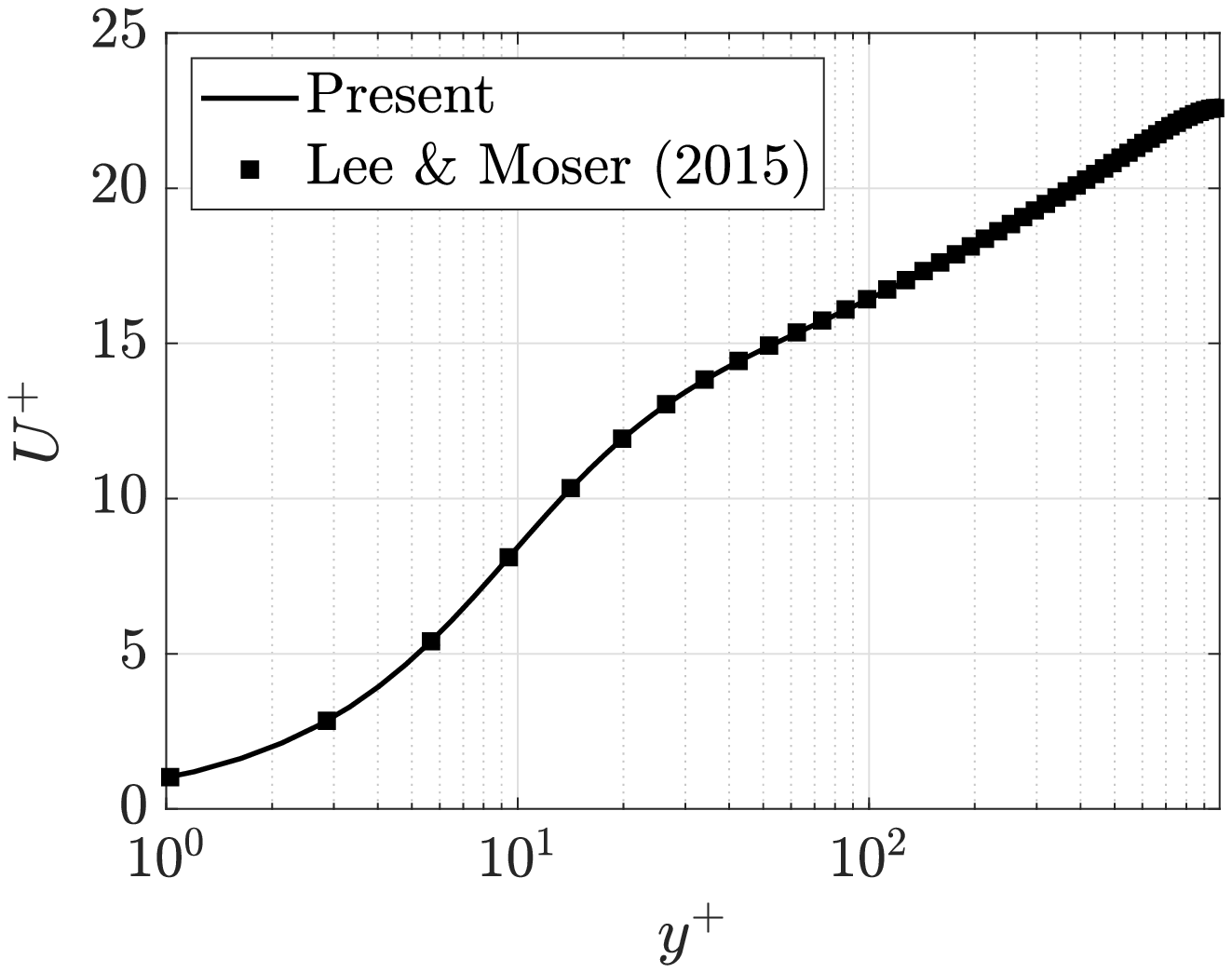}
		\caption{}
		\label{fig:Retau1000_validation_uplus_profile}
	\end{subfigure}
	\begin{subfigure}{0.44\textwidth}
		\includegraphics[width=\textwidth]{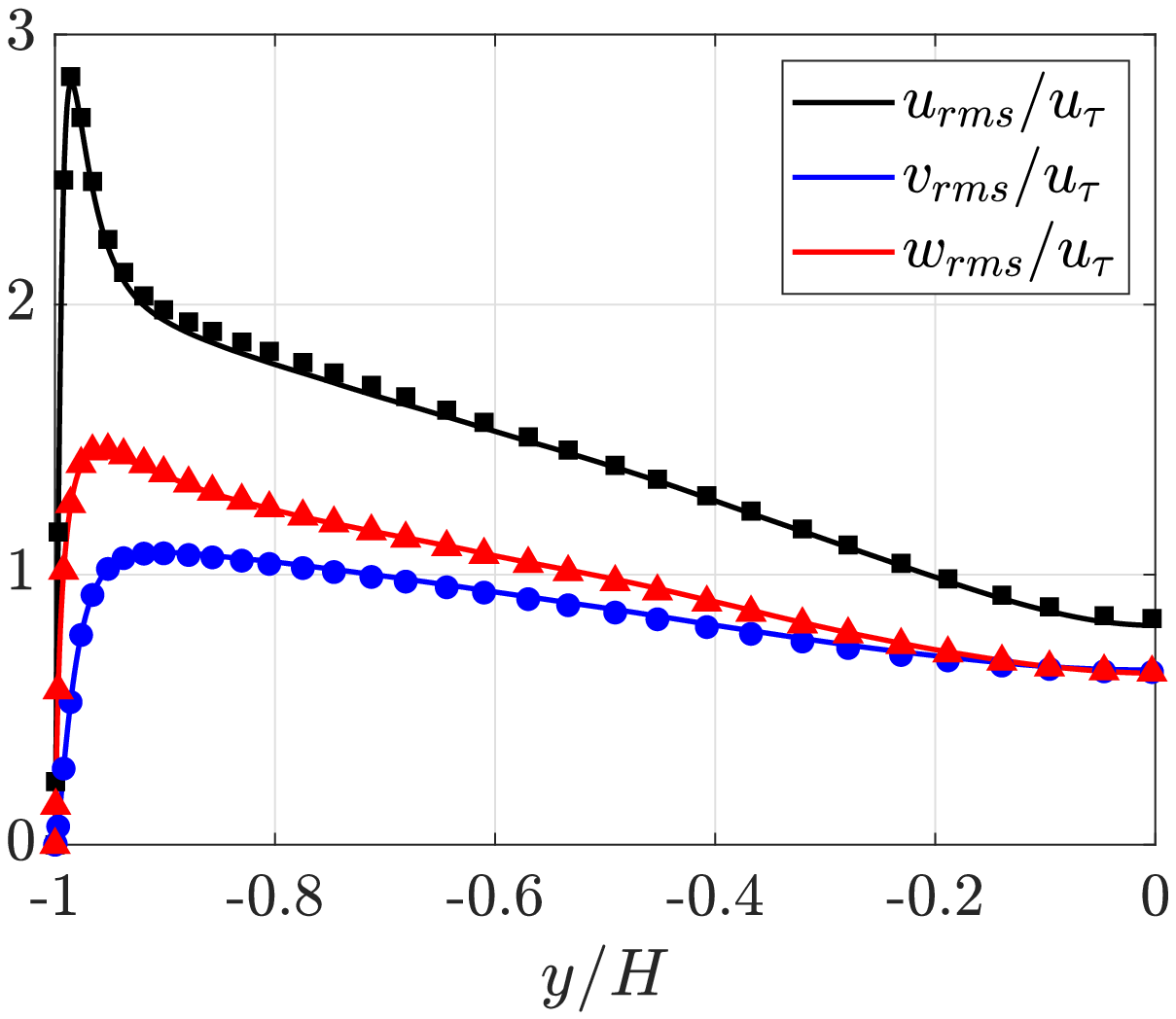}
		\caption{}
		\label{fig:Retau1000_validation_rms}
	\end{subfigure}
	\caption{$Re_{\tau} \approx 1000$ channel flow DNS validation results. Lines indicate the present simulations results and symbols denote data from \citet{lee2015direct}. Left frame: mean velocity profile ($U^{+}$). Right frame: root-mean-square (RMS) values for each velocity component.}
	\label{fig:Retau1000_validation}
\end{figure}

The DNS databases for $Re_{\tau} \approx 180$ and 550 were validated by \citet{martini2020resolvent} and \citet{morra2021colour}, respectively, using earlier simulations by \citet{delalamo2003spectra}.
We validate the $Re_{\tau} \approx 1000$ database against results from \citet{lee2015direct} in figure \ref{fig:Retau1000_validation}.
In the figure, $H$ denotes the channel half-height, $y^+$ is the wall-normal position in inner (viscous) units, $U^+$ is the mean velocity profile in inner units, $u$, $v$ and $w$ correspond to the streamwise, wall-normal and spanwise velocity components, subscript $rms$ indicates root-mean-square values and $u_{\tau}$ is the friction velocity.
Good agreement is observed, validating the present database.

Table \ref{tab:dns_parameters} shows, among other information, the Reynolds numbers (based on the friction velocity $u_{\tau}$, i.e., $Re_{\tau}$, and channel half-height $H$ and bulk velocity $U_b$, i.e., $Re$), number of mesh points ($N_x$, $N_y$ and $N_z$) and mesh discretisation ($\Delta x^{+}$, $\Delta z^{+}$, $\Delta {y_{min}}^{+}$ and $\Delta {y_{max}}^{+}$).
In all cases, a plus superscript indicates non-dimensional quantities using inner (viscous) scaling.
Quantities expressed in outer units are denoted without the plus superscript and are normalised by the bulk velocity and channel half height.
The number of wavenumbers in the streamwise and spanwise directions in the DNS is $3/2$ times higher than the $N_x$ and $N_z$ values reported in Table \ref{tab:dns_parameters} for de-aliasing.
The number of snapshots ($N_t$) and the time steps based on inner units ($\Delta t^{+}$) are also exhibited in the table.
In the simulations, initial transients, detected by evaluating the turbulent kinetic energy and the friction coefficient on the walls, were discarded; thus the snapshots retained are within the statistical steady state.
The number of time-steps ($N_{fft}$) in each block of data and the number of blocks ($N_b$) used within Welch's method \citep{welch1967fft} to estimate the CSDs and power spectral densities (PSDs) of DNS and estimation data are also provided in Table \ref{tab:dns_parameters}.
A Hann window and 75\% overlap was used to compute spectra.

\begin{table}
  \begin{center}
		\def~{\hphantom{0}}
		\begin{tabular}{c c c c c c c c c c c c c}
			$Re_{\tau}$	& $Re$  & $N_x$	& $N_y$	& $N_z$	& $\Delta x^{+}$	& $\Delta z^{+}$	& $\Delta {y_{min}}^{+}$	& $\Delta {y_{max}}^{+}$	& $N_t$	& $\Delta t^{+}$	& $N_{fft}$	& $N_b$	\\ [3pt]
			180 (179)		& 2800	& 192		& 129		& 192		& 11.71						& 5.86						& 5.39 $\times$ 10$^{-2}$	& 4.39										& 4797	& 5.72						& 256				& 34	\\
			550 (543)		& 10000 & 384 	& 257		& 384		& 8.89						& 4.44						& 4.09 $\times$ 10$^{-2}$	& 6.66										& 3000	& 2.95						& 256				& 43	\\
			1000 (996)	& 20000	& 484		& 385		& 484		& 12.93						& 6.47						& 3.33 $\times$ 10$^{-2}$	&	8.15										& 3961	& 2.48						& 512				& 27	\\
		\end{tabular}
		\caption{Channel flow DNS parameters for the cases studied.}
		\label{tab:dns_parameters}
  \end{center}
\end{table}

In order to reduce computational cost and storage requirements, the DNS data (state and forcing terms) were then filtered, retaining a lower number of wavenumbers in the streamwise and spanwise directions, e.g., $N_{\alpha} = 31$ and $N_{\beta} = 32$ for the $Re_{\tau} \approx 550$ case (see Table \ref{tab:cutoff_parameters} and App. \ref{app:DNS_filtering}).
Following this strategy, it was possible to compress the data, reducing the storage and memory needed to perform, \textit{a posteriori}, the resolvent-based estimations.
As an example, for $Re_{\tau} \approx 550$, a total of $N_x$ and $N_z$ discrete $\alpha$ and $\beta$, respectively, are possible for the current mesh parameters, i.e., the database was compressed by a factor of approximately 148.
Note that we make use of the property $\boldsymbol{q}(\alpha,y,\beta,t) = \boldsymbol{q}^{*}(-\alpha,y,-\beta,t)$.
The cutoff values for $\alpha$ and $\beta$ were defined based on the premultiplied power-spectral peak at a wall-normal distance of $y^{+} = 15$, i.e., $({\lambda_x}^{+},~{\lambda_z}^{+}) \approx (1000,~100)$ \citep{delalamo2003spectra}, where ${\lambda_x}^{+}$ and ${\lambda_z}^{+}$ are the streamwise and spanwise wavelengths in wall units.
The $\alpha$ and $\beta$ cutoff values were chosen such that all the cases include the wavenumber corresponding to $({\lambda_x}^{+},~{\lambda_z}^{+}) \approx (1000,~100)$, ensuring that the premultiplied power spectra peak is contained within the data.
The smaller flow structures are filtered as a consequence of data compression, but this does not impact the estimation of the retained wavenumbers due to the linearity of the estimator.
Table \ref{tab:cutoff_parameters} provides the precise cutoff values in outer ($\alpha_{cut}$ and $\beta_{cut}$) an inner (${\alpha_{cut}}^{+}$ and ${\beta_{cut}}^{+}$) units.

\begin{table}
  \begin{center}
		\def~{\hphantom{0}}
		\begin{tabular}{c c c c c c c c}
			$Re_{\tau}$	& $N_{\alpha}$	& $N_{\beta}$	& $\alpha_{cut}$	& $\beta_{cut}$	& ${\alpha_{cut}}^{+}$	& ${\beta_{cut}}^{+}$	& Downsampling ratio	\\ [3pt]
			180 (179)		& 15						& 32					& 8								& 31						&	0.0447								&	0.1732							& 77								\\
			550 (543)		& 31						& 32					& 16							& 62						&	0.0295								& 0.1142							& 148								\\
			1000 (996)	& 31						& 40					& 15							& 78						&	0.0151								&	0.0783							& 188								\\
		\end{tabular}
		\caption{Wavenumber cutoff values used to filter the DNS databases.}
		\label{tab:cutoff_parameters}
  \end{center}
\end{table}

\section{Case study: $Re_{\tau} \approx$ 550}
\label{sec:Retau550}

In this section, a comprehensive analysis of the $Re_{\tau} \approx 550$ estimates is conducted.
Three models for the forcing terms, i.e., the non-linearities on the Navier-Stokes system, are addressed, including the true forcing statistics (obtained from the DNS), white-noise forcing and an implicitly coloured forcing obtained by including an eddy viscosity in the linearised operator.
Different choices of sensors on the channel walls are also explored.
Several types of comparison with the filtered DNS data are used to assess the estimates, including reconstructions of sample snapshots, RMS error, correlations, and power spectra.

\subsection{Estimation of a sample snapshot}
\label{sec:Retau550_snapshots}

\begin{figure}
	\centerline{\includegraphics[width=\textwidth]{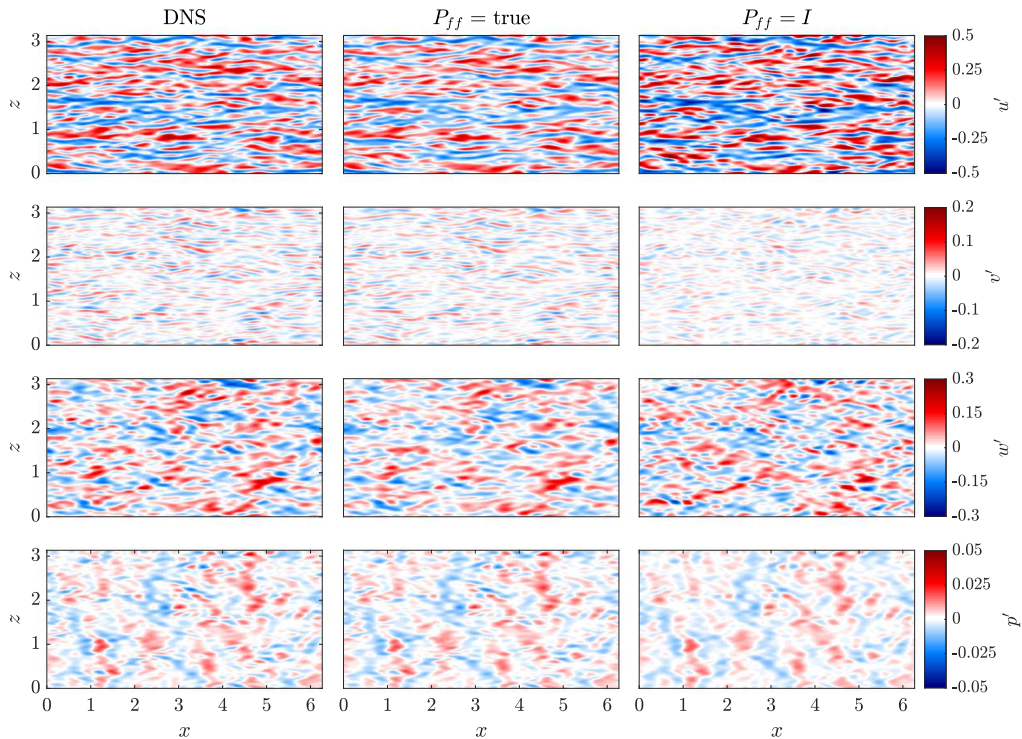}}
	\caption{Comparison between filtered DNS and resolvent-based estimates of an instantaneous snapshot of the flow state for $Re_{\tau} \approx 550$, $y^{+} \approx 15$ and using wall measurements of pressure and shear stress. Columns, from left to right: DNS data, true and white noise forcing estimates. Rows, from top to bottom: streamwise ($u^{\prime}$), wall-normal ($v^{\prime}$), spanwise ($w^{\prime}$) velocity fluctuations, respectively, and pressure fluctuation ($p^{\prime}$). Fluctuations shown in outer units. An animated version of this figure showing its time evolution is provided as supplementary material.}
	\label{fig:q_estimation_Retau550_yplus15}
\end{figure}

\begin{figure}
	\centerline{\includegraphics[width=\textwidth]{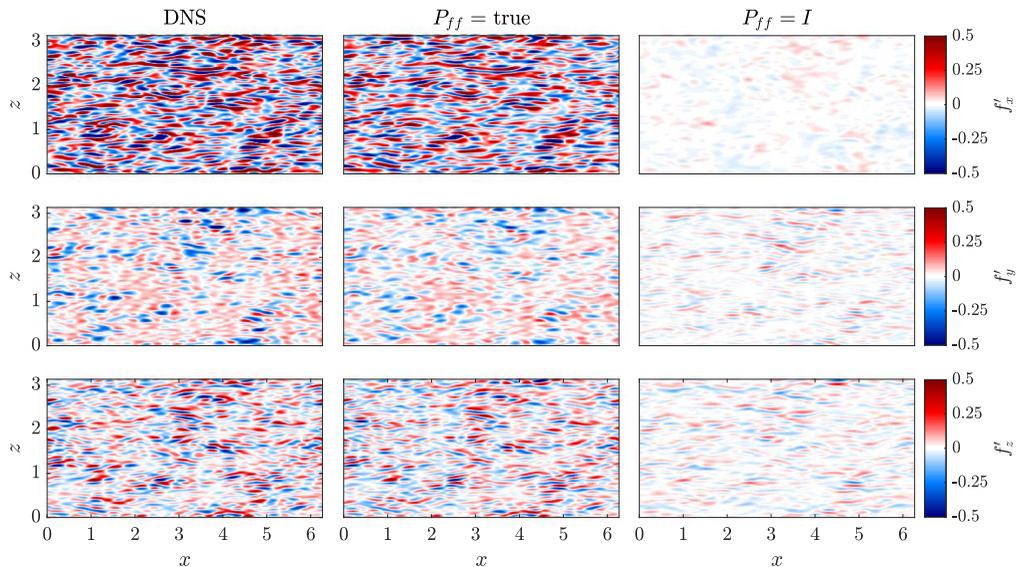}}
	\caption{Comparison between filtered DNS and resolvent-based estimates of an instantaneous snapshot of the flow forcing components for $Re_{\tau} \approx 550$, $y^{+} \approx 15$ and using wall measurements of pressure and shear stress. Columns, from left to right: DNS data, true forcing statistics and white noise forcing estimates. Rows, from top to bottom: streamwise ($f_{x}^{\prime}$), wall-normal ($f_{y}^{\prime}$) and spanwise ($f_{z}^{\prime}$) forcing fluctuation components, respectively. Fluctuations shown in outer units.}
	\label{fig:f_estimation_Retau550_yplus15}
\end{figure}

\begin{figure}
	\centerline{\includegraphics[width=\textwidth]{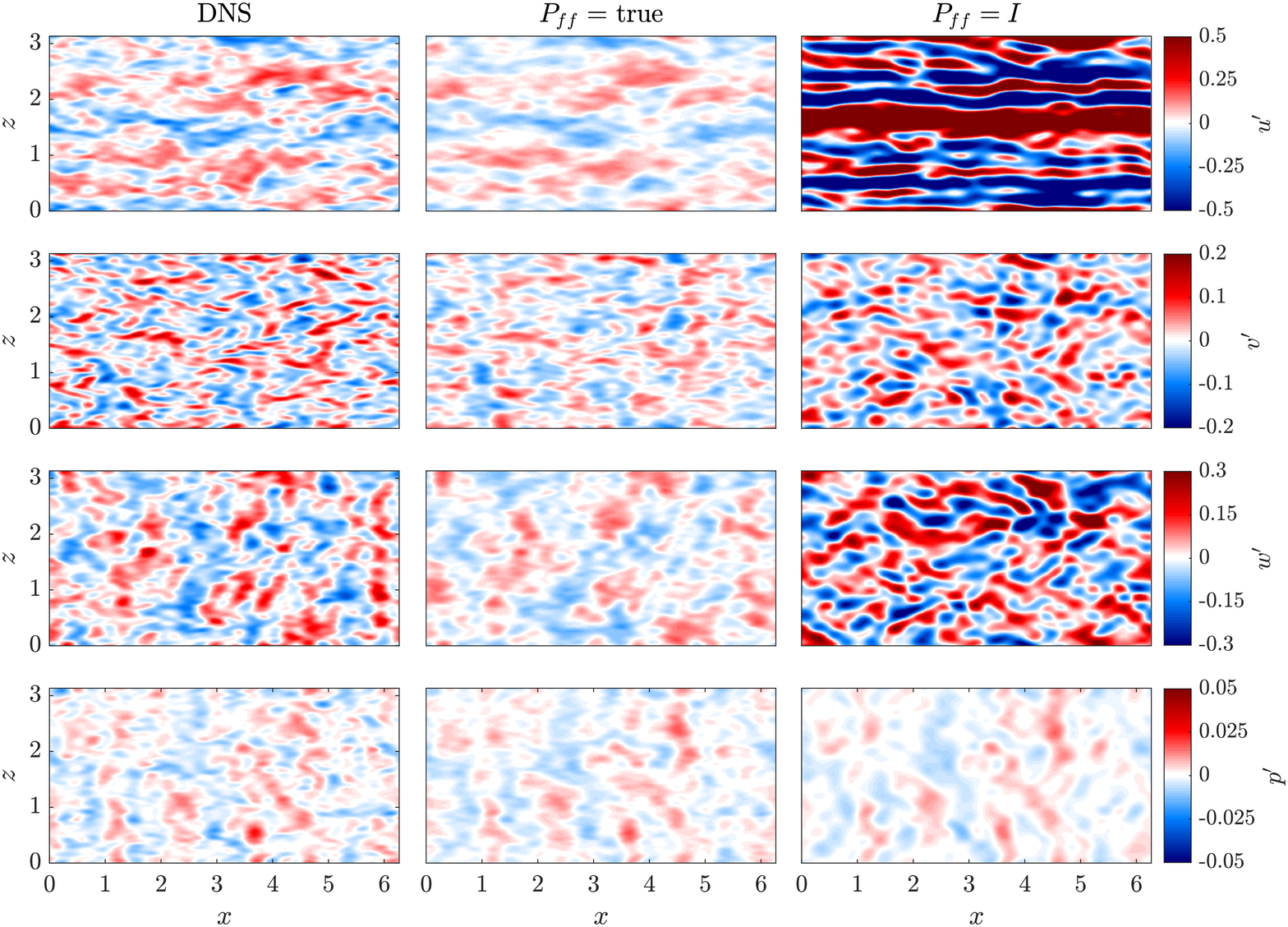}}
	\caption{Comparison between filtered DNS and resolvent-based estimates of an instantaneous snapshot of the flow state for $Re_{\tau} \approx 550$, $y^{+} \approx 100$ and wall measurements of pressure and shear stress. See comments in the caption of figure \ref{fig:q_estimation_Retau550_yplus15}.}
	\label{fig:q_estimation_Retau550_yplus100}
\end{figure}

\begin{figure}
	\centerline{\includegraphics[width=\textwidth]{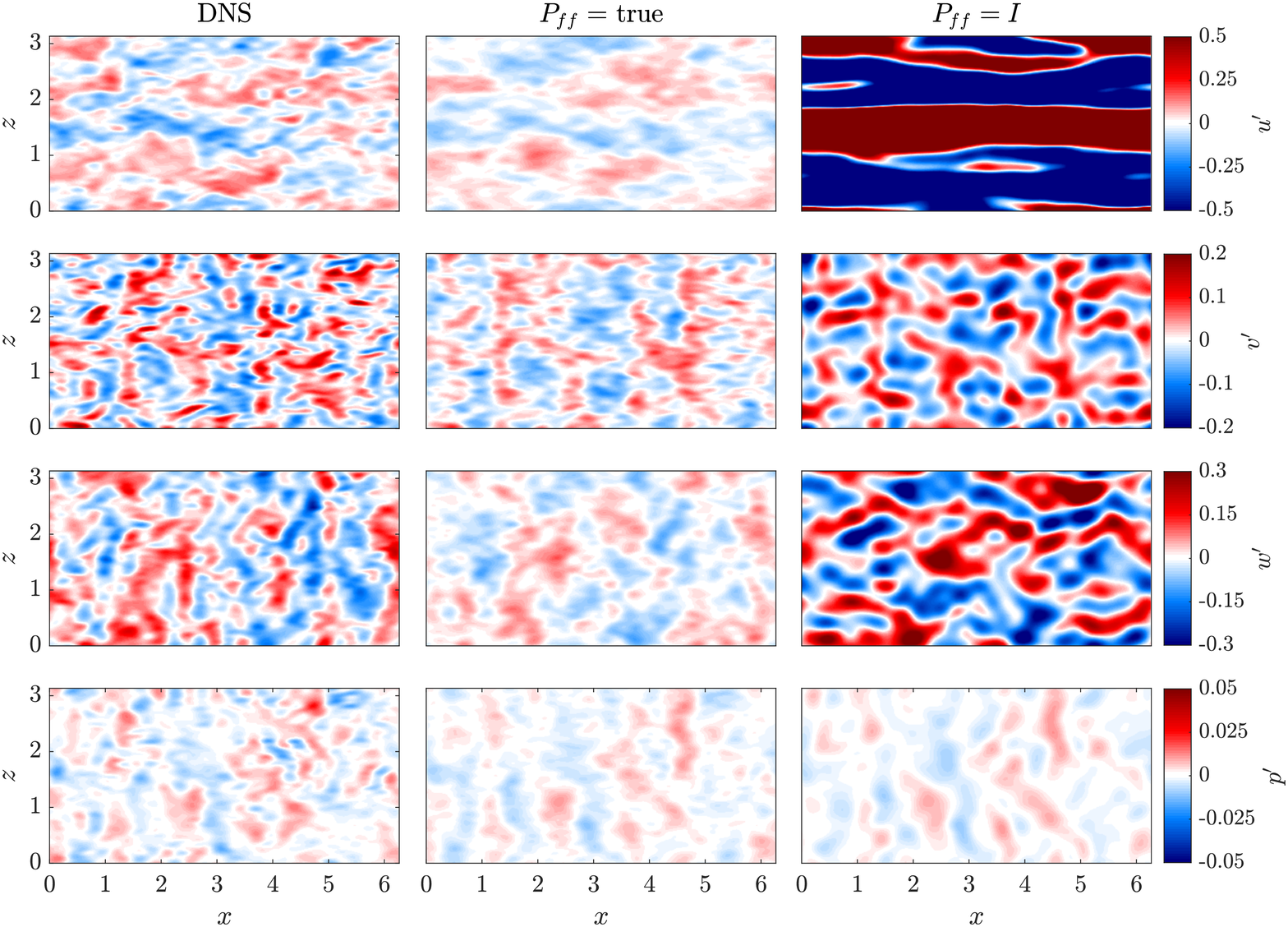}}
	\caption{Comparison between filtered DNS and resolvent-based estimates of an instantaneous snapshot of the flow state for $Re_{\tau} \approx 550$, $y^{+} \approx 200$ and wall measurements of pressure and shear stress. See comments in the caption of figure \ref{fig:q_estimation_Retau550_yplus15}.}
	\label{fig:q_estimation_Retau550_yplus200}
\end{figure}

To provide a first observation of the estimated structures in the turbulent field, we begin by comparing flow estimate and DNS results for sample snapshots, taken at some reference wall-parallel planes.
Filtered DNS snapshots, retaining only the wavenumbers considered in the estimations, are displayed in the figures to enable direct comparisons; appendix \ref{app:DNS_filtering} shows DNS snapshots without the application of the spatial filtering for the state and forcing components.
Snapshots of flow-field state and forcing estimates are shown in figures \ref{fig:q_estimation_Retau550_yplus15} through \ref{fig:q_estimation_Retau550_yplus200}.
Figures \ref{fig:q_estimation_Retau550_yplus15} and \ref{fig:f_estimation_Retau550_yplus15} show, respectively, state and forcing estimates in the buffer layer, at $y^{+} \approx 15$, and figures \ref{fig:q_estimation_Retau550_yplus100} and \ref{fig:q_estimation_Retau550_yplus200} show state estimates at $y^{+} \approx 100$ and 200.
In all cases, estimates from wall measurements are compared to the DNS results.
When the true forcing statistics are considered ($\mathsfbi{P_{ff}} = \mathsfbi{P_{ff,DNS}}$), the resemblance is remarkable between the filtered DNS and the resolvent-based estimates of all flow state components in the buffer layer, at a wall-normal distance of $y^{+} \approx 15$ (figure \ref{fig:q_estimation_Retau550_yplus15}).
Even when the forcing CSD is approximated as white noise ($\mathsfbi{P_{ff}} = \mathsfbi{I}$), close agreement between the DNS and estimated states is observed for $y^{+} \approx 15$.
Both estimates were able to recover the near-wall streaks of streamwise velocity $u$, as well as the structures for other velocity components and pressure.

Figure \ref{fig:f_estimation_Retau550_yplus15} shows snapshots of the forcing components.
Good agreement between filtered DNS data and estimates is achieved, i.e., most of the structures present on the DNS were captured by the estimates.
Note that although the true colour was employed as the forcing CSD ansatz in (\ref{eq:Tf}) to obtain the results displayed in the figure, the estimated forcing components were not exactly recovered, as the system observation was restricted to the wall measurements of shear stresses and pressure.
Forcing estimates obtained with the white-noise forcing ansatz have a significant mismatch with the DNS reference values.

Figures \ref{fig:q_estimation_Retau550_yplus100} and \ref{fig:q_estimation_Retau550_yplus200} show the state-estimate snapshots in the logarithmic layer, at $y^{+} \approx 100$ and $\approx 200$, respectively, for $Re_{\tau} \approx 550$.
These are more challenging estimates, as they involve positions farther from the wall, where measurements are taken.
The estimates were not as accurate as the previous results obtained for $y^{+} \approx 15$ (figure \ref{fig:q_estimation_Retau550_yplus15}).
Nevertheless, most of the large-scale structures present in the DNS snapshots were well captured by the estimation procedure, even for $y^{+} \approx 200$.
Far from the wall, the white noise forcing assumption does not lead to results of the same quality as the those obtained close to the wall, figure \ref{fig:q_estimation_Retau550_yplus15}.
Such results are in line with recent works \citep{illingworth2018estimating, abreu2020resolvent, morra2021colour}, which have shown that considering the forcing CSD to be white noise in a resolvent formulation leads to reasonable models for near-wall structures but significant mismatch for larger-scale structures at higher wall-normal positions.

Most structures within the velocity field are faithfully recovered by the estimates using the true forcing statistics, albeit with somewhat lower amplitude than the DNS fields in many cases.
On the other hand, when white-noise forcing is considered, shown in the right columns, amplitudes were overestimated.
Only the pressure component could be estimated with reasonable accuracy using the white noise forcing assumption.
This is likely related to the non-local behaviour of pressure fluctuations, which are related to the global velocity fluctuations by a Poisson equation \citep{anantharamu2020analysis}; thus, pressure fluctuations tend to be more spatially extended, simplifying their estimation from wall measurements.
Forcing-component estimates are not displayed, as the estimates were less accurate than those shown in figure \ref{fig:f_estimation_Retau550_yplus15}, especially regarding their amplitudes.

\subsection{Effect of the choice of sensors}
\label{sec:Retau550_sensors}

\begin{figure}
	\centerline{\includegraphics[width=\textwidth]{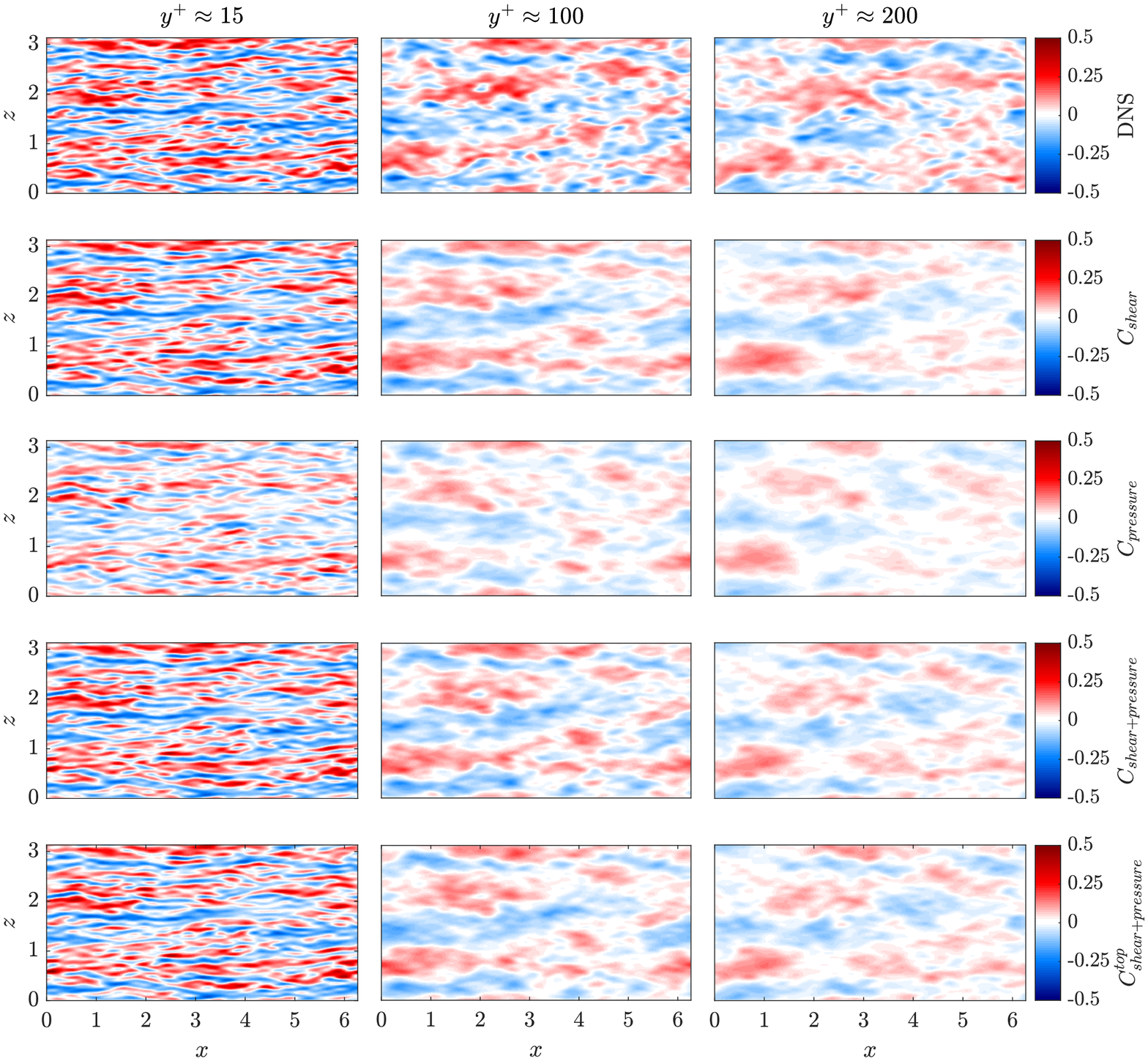}}
	\caption{Comparison among DNS and true forcing statistics of the streamwise velocity estimates obtained with different definitions for the observation matrix $\mathsfbi{C}$ for $Re_{\tau} \approx 550$. Columns, from left to right: $y^{+} \approx 15$, 100 and 200. Rows, from top to bottom: DNS, observation $\mathsfbi{C}$ containing only the wall shear stress ($\frac{du}{dy}\rvert_{wall}$ and $\frac{dw}{dy}\rvert_{wall}$) on both walls, observation $\mathsfbi{C}$ containing only the wall pressure ($p_{wall}$) on both walls, observation $\mathsfbi{C}$ containing both wall shear stress and pressure ($\frac{du}{dy}\rvert_{wall}$, $\frac{dw}{dy}\rvert_{wall}$ and $p_{wall}$) on both walls and observation $\mathsfbi{C}$ containing both wall shear stress and pressure ($\frac{du}{dy}\rvert_{wall}$, $\frac{dw}{dy}\rvert_{wall}$ and $p_{wall}$) only on the top wall. Fluctuations shown in outer units. An animated version of this figure is provided as supplementary material.}
	\label{fig:u_C_Retau550}
\end{figure}

Figure \ref{fig:u_C_Retau550} displays snapshots of the streamwise velocity component from the DNS and estimates obtained with different definitions for the observation matrix $\mathsfbi{C}$, i.e., sensors accounting for the wall shear stress only ($\frac{du}{dy}\rvert_{wall}$ and $\frac{dw}{dy}\rvert_{wall}$), wall pressure only ($p_{wall}$) and both wall shear stress and pressure.
The measurements were taken at both top and bottom channel walls (second to fourth rows) and only on the channel top wall (fifth row), for the $Re_{\tau} \approx 550$ case and three wall-normal positions, i.e., $y^{+} \approx 15$, 100 and 200, which are shown in the first, second and third columns of the figure, respectively.
To compare the different estimates, true forcing statistics were used.

The estimates are less accurate when only observations of the pressure on both walls are considered.
When both wall shear stress and pressure are provided to evaluate the transfer function between the measurements and estimated state, better results are achieved.
The case where only the wall shear stress is used displays an intermediate performance, only slightly worse than estimates from wall shear stress plus wall pressure.
Using both shear stress and pressure at just one wall (the one closer to the wall-normal planes where the results are evaluated) leads to estimates of similar quality as those obtained with measurements at both walls.
Appendix \ref{app:Cmatrix_errors} shows comparison metrics for the different observation definitions to corroborate these conclusions.

\subsection{Estimation with the linearised operator including eddy-viscosity}
\label{sec:Retau550_EV}

\begin{figure}
	\centerline{\includegraphics[width=\textwidth]{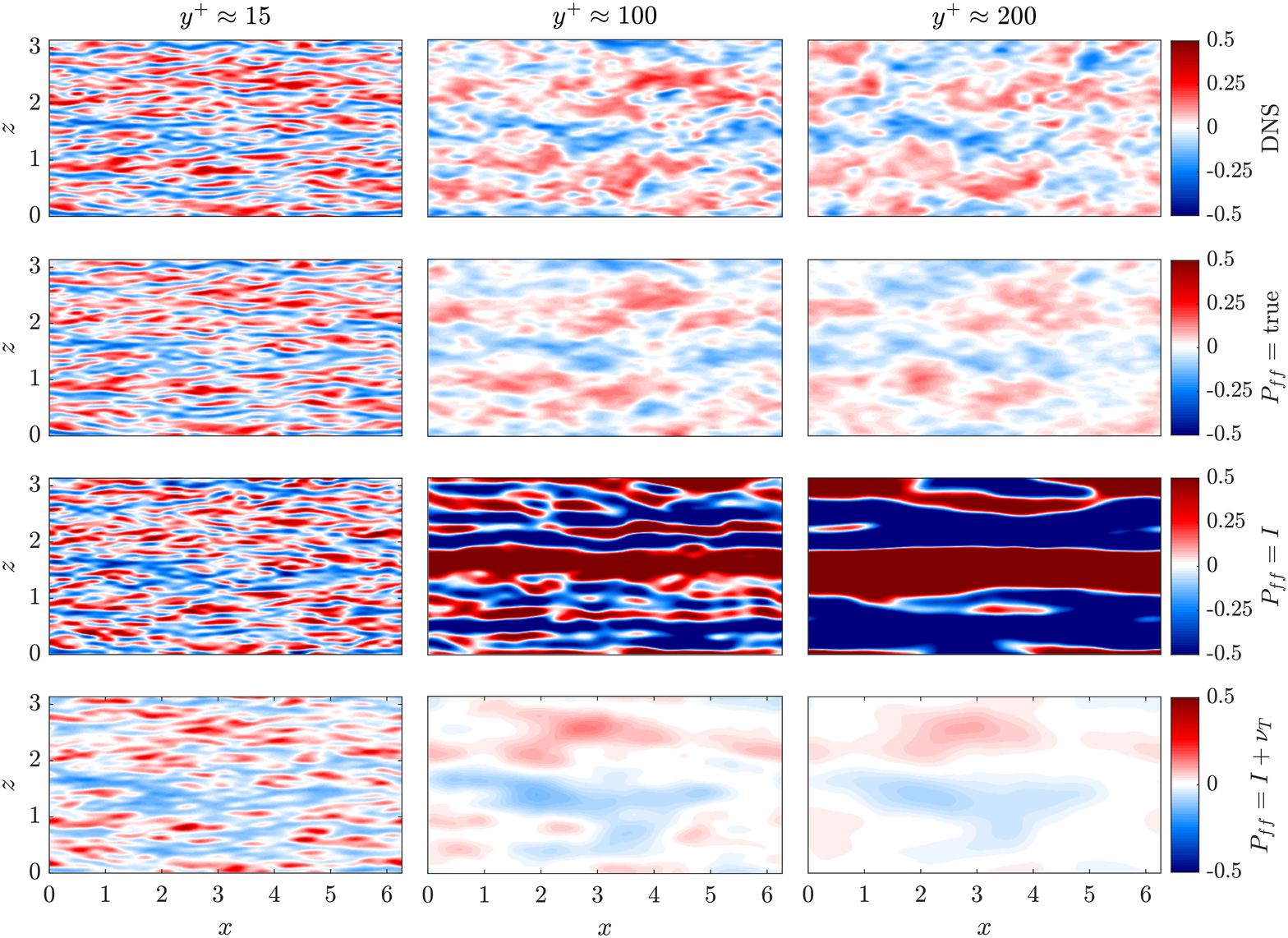}}
	\caption{Comparison between DNS and estimates considering the eddy-viscosity model. Streamwise velocity component ($u^{\prime}$) snapshots for $Re_{\tau} \approx 550$. Columns, from left to right: $y^{+} \approx 15$, 100 and 200. Rows, from top to bottom: DNS data and estimates using the true, white noise and eddy-viscosity forcing models. Fluctuations shown in outer units. An animated version of this figure is provided as supplementary material.}
	\label{fig:u_nuT_Retau550}
\end{figure}

Earlier works have obtained flow estimates using a linearised operator including eddy-viscosity \citep{illingworth2018estimating, towne2020resolvent} and white noise forcing.
Such a model attempts to partially represent the colour of the non-linear terms of the Navier-Stokes system, which improves predictions obtained from the resolvent framework \citep{morra2019relevance}.
To make comparisons with this approach, we employ the same methodology used by \citet{towne2020resolvent}, which considers the total viscosity $\nu_T$; appendix \ref{app:math} shows the linear operator formulation.
This approach will be referred to as the eddy-viscosity model.
Figure \ref{fig:u_nuT_Retau550} exhibits the estimates for the $Re_{\tau} \approx 550$ case and three wall-normal positions, similar to figure \ref{fig:u_C_Retau550}, but with the rows representing the DNS results and true, white and eddy-viscosity model ($\nu_T$) estimates.
Only streamwise velocity component results are displayed in the figure, and similar snapshot reconstructions were obtained for the other state components.

The use of the eddy-viscosity model prevented the large amplitude overprediction in the estimates observed in the white noise forcing case and is thus an option when the forcing statistics are unknown.
At the same time, the estimates obtained using the eddy-viscosity model are significantly worse than than those obtained using the true forcing statistics from the DNS.
Another aspect worthy of mention is the eddy-viscosity model has a dual effect on the estimations compared to the white noise forcing statistics: while it improves estimates of large-scale coherent structures further from the wall ($y^{+} \approx 100$ and 200 snapshots), it worsens the estimates of near-wall structures ($y^{+} \approx 15$).
Improved results could be obtained if, instead of using the standard Cess model (see Appendix \ref{app:math}), the eddy-viscosity model was to be calibrated for each friction Reynolds number \citep{pickering2021optimal}.
However, our goal is not to find the optimal viscosity model for the flow, but rather to show that these models can improve the estimation by acting as a surrogate for the true, but unknown, force colour.

Several studies with linearised models showed better agreement with turbulence statistics when an eddy viscosity is included in the linearised operator \citep{del2006linear, hwang2010linear, morra2019relevance, pickering2021optimal}.
As discussed by \citet{morra2021colour}, the inclusion of an eddy viscosity may be seen as the introduction of some forcing colour to the linearised operator based on the molecular viscosity.
Although this is not exact, this modelling approach is found to improve predictions from the linearised system, particularly for large-scale structures, as seen in the present estimates using white-noise forcing (bottom two rows of figure \ref{fig:u_nuT_Retau550}).
However, despite these improvements, the consideration of white-noise forcing in an operator that includes an eddy viscosity is still an approximation that leads to errors for a number of structures, as recently discussed by \citet{symon2021energy}.
Use of the actual statistics of non-linear terms from a simulation leads to an exact agreement in the input-output system by construction \citep{towne2018spectral, morra2021colour}, and this is, therefore, the best-case scenario for a linearised estimator.
This is confirmed by the superior performance of estimates using $\mathsfbi{P_{ff}} = \mathsfbi{P_{ff,DNS}}$ compared to the other modelling strategies.
However, this comes with the additional requirement of forcing statistics, leading to a more complex procedure, as discussed in \S \ref{sec:Retau550_forcing modelling}.

\subsection{Statistics of the estimated flow}
\label{sec:Retau550_statistics}

\begin{figure}
	\centerline{\includegraphics[width=\textwidth]{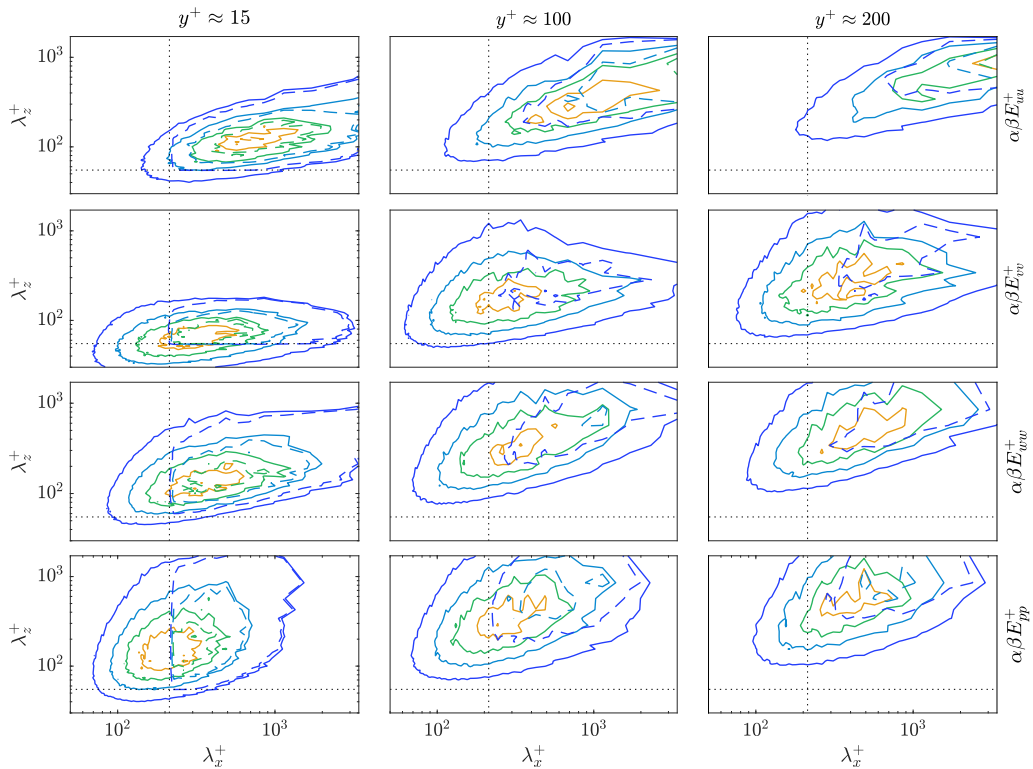}}
	\caption{Premultiplied power spectra comparison between DNS (solid lines) and resolvent-based estimates considering true forcing statistics (dashed lines) and wall measurements of pressure and shear stress for $Re_{\tau} \approx 550$. Columns, from left to right: $y^{+} \approx 15$, 100 and 200. Rows, from top to bottom: streamwise ($\alpha \beta {E_{uu}}^+$), wall-normal ($\alpha \beta {E_{vv}}^+$) and spanwise ($\alpha \beta {E_{ww}}^+$) velocity and pressure ($\alpha \beta {E_{pp}}^+$) premultiplied power spectra. Contour levels equally spaced between 0.2 and 0.8 times the peak value of the DNS. Dotted lines indicate the wavenumbers cutoff quadrant.}
	\label{fig:pre-multiplied_energy_spectra_Retau550_PffTrue}
\end{figure}

\begin{figure}
	\centerline{\includegraphics[width=\textwidth]{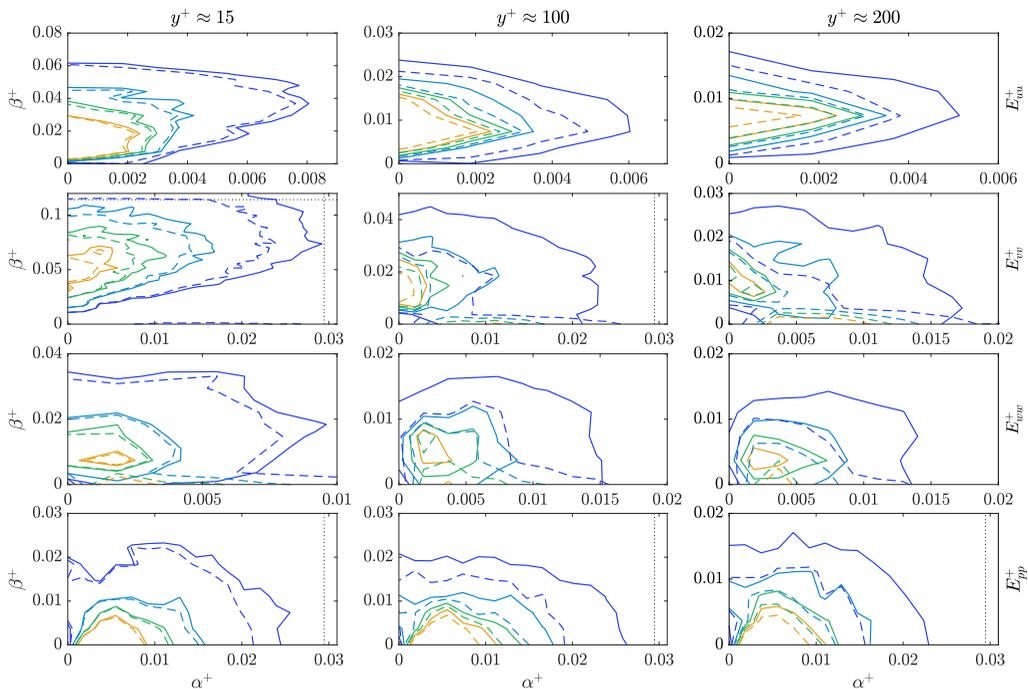}}
	\caption{Power spectra comparison between DNS (solid lines) and resolvent-based estimates using true forcing statistics (dashed lines) and wall measurements of pressure and shear stress for $Re_{\tau} \approx 550$. Columns, from left to right: $y^{+} \approx 15$, 100 and 200. Rows, from top to bottom: streamwise (${E_{uu}}^+$), wall-normal (${E_{vv}}^+$) and spanwise (${E_{ww}}^+$) velocity and pressure (${E_{pp}}^+$) power spectra. See comments in the caption of figure \ref{fig:pre-multiplied_energy_spectra_Retau550_PffTrue}.}
	\label{fig:energy_spectra_Retau550_PffTrue}
\end{figure}

Figures \ref{fig:pre-multiplied_energy_spectra_Retau550_PffTrue} and \ref{fig:energy_spectra_Retau550_PffTrue} show the power spectra of velocity components with and without premultiplication, respectively, for different wall-normal distances.
Premultiplication of spectra by $\alpha$ and $\beta$ highlights smaller-scale fluctuations, which are not accurately estimated from wall measurements as seen in figures \ref{fig:q_estimation_Retau550_yplus100} and \ref{fig:q_estimation_Retau550_yplus200}; hence, we have also included spectra without premultiplication in figure \ref{fig:energy_spectra_Retau550_PffTrue} for a clearer visualization of the energy content of the largest scales in the simulation.
The estimates were evaluated using the true forcing statistics.
Continuous and dashed lines represent the power spectra computed from the full DNS snapshots (without filtering) and the resolvent-based estimates, respectively.
Four equally spaced contours levels are plotted and all levels are normalised by the peak value of the DNS spectra for each state component and wall-normal distance.
The dotted lines indicate the wavenumber cutoff; only spectral values in the first quadrant, formed by the intersection of these crossing lines, i.e., the top-right portion of each frame, are estimated.

The agreement between the DNS and estimates is excellent for all state components at a wall-normal distance of $y^{+} \approx 15$, consistent with the observations of figure \ref{fig:q_estimation_Retau550_yplus15}.
The peak in the premultiplied power spectrum peak at $({\lambda_x}^{+},~{\lambda_z}^{+}) \approx (1000,~100)$ is accurately estimated for the streamwise velocity component at $y^{+} \approx 15$.
At this wall-normal distance, the peak energies of the other state components are also captured by the estimates.
However, for the pressure component, only a portion of the peak was estimated as the bulk containing the most energetic part of the spectrum lies outside the observable wavenumber parameters.
Further from the wall, the estimated spectra progressively deteriorate.
This effect appears stronger in the premultiplied spectra of figure \ref{fig:pre-multiplied_energy_spectra_Retau550_PffTrue}, which highlights smaller structures, while spectra without premultiplication, which highlights larger scales, show a better agreement, with the estimation leading to some underestimation of the peak power.

\begin{figure}
	\centerline{\includegraphics[width=\textwidth]{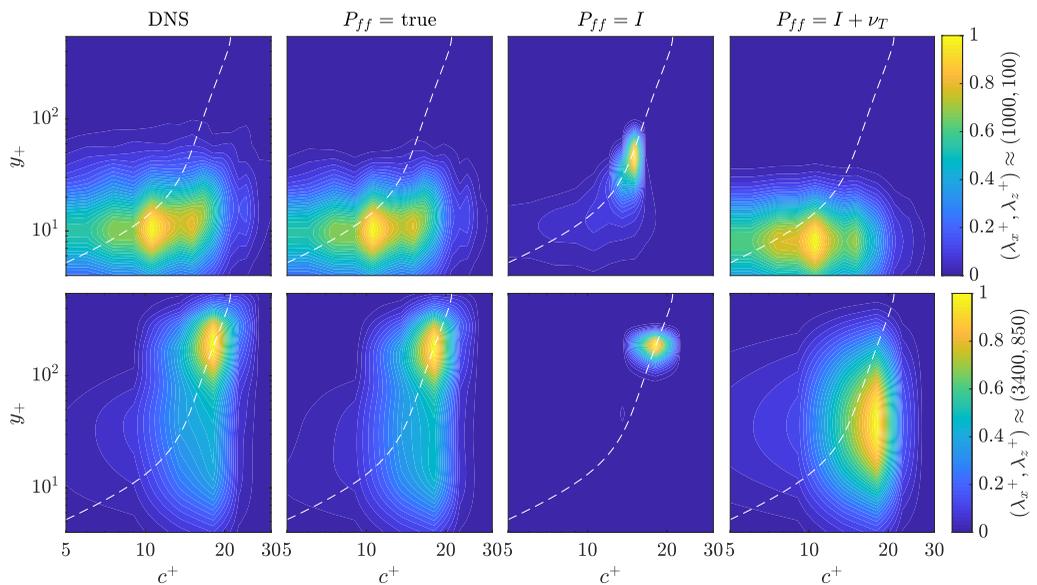}}
	\caption{Comparison between DNS and resolvent-based estimates of the streamwise velocity PSDs for $Re_{\tau} \approx 550$ using wall measurements of pressure and shear stress. PSD contour plots are normalised by the peak value at each frame. The dashed lines indicate the mean velocity profile in plus units ($U^{+}$). Columns, from left to right: DNS data, true, white noise and eddy-viscosity model estimates. Top row: $({\lambda_x}^{+},~{\lambda_z}^{+}) \approx (1000,~100)$ (near-wall structures). Bottom row: $({\lambda_x}^{+},~{\lambda_z}^{+}) \approx (3400,~850)$ (large-scale structures).}
	\label{fig:PSD_Retau550}
\end{figure}

Figure \ref{fig:PSD_Retau550} shows the PSD of the streamwise velocity as a function of the wall-normal coordinate ($y^{+}$) and phase speed $c^{+} = \omega^{+}/\alpha^{+}$ for two combinations of $({\lambda_x}^{+},~{\lambda_z}^{+})$, i.e., $\approx (1000,~100)$ and $(3400,~850)$.
Such wavenumber combinations correspond to the near-wall and large-scale structures respectively, extracted from peaks of premultiplied spectra for the $Re_{\tau} \approx 550$ case \citep{morra2021colour}.
In the figure, the dashed lines represent the mean velocity profile in plus units.
The position of peak PSD $(y^{+}, c^{+})$ is related to the critical layer ($U^{+} = c^{+}$), as expected from an analysis of the resolvent operator \citep{mckeon2010critical, tissot2017sensitivity}.
The white-noise forcing assumption is only able to estimate the PSD peak, which follows the critical layer, whereas the estimates obtained using the true forcing statistics are closer to the DNS results.
This shows that incorporating the forcing colour produces a model that is able to capture the bulk of the energy that lies outside the critical\textcolor{blue}{-}layer position.
When the eddy-viscosity model is considered, the white\textcolor{blue}{-}noise forcing results are improved, especially for the wavenumber pair related to the near-wall structures, but errors are nonetheless present for larger structures.

\begin{figure}
	\centerline{\includegraphics[width=\textwidth]{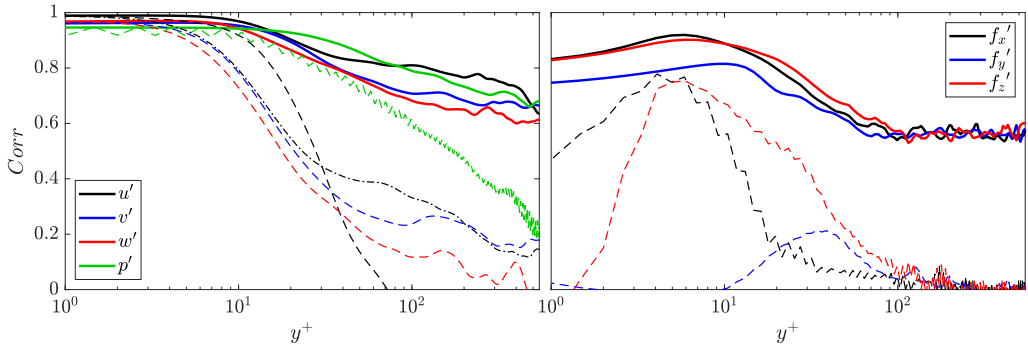}}
	\caption{Correlation between filtered DNS and estimate data for $Re_{\tau} \approx 550$ case and wall measurements of shear stress and pressure. The left and right frames show the flow state and forcing components, respectively. Continuous, dashed and dash-dotted lines denote estimates obtained with true forcing statistics, white-noise and the eddy-viscosity model, respectively.}
	\label{fig:qf_correlation_Retau550}
\end{figure}

Normalised correlations between the filtered DNS data and resolvent-based estimates are shown in figure \ref{fig:qf_correlation_Retau550}.
The correlation metric is defined as
\begin{equation}
	Corr(x,y,z) = \frac{\int {q_{DNS}}\left(x,y,z,t\right) {q_{est}}\left(x,y,z,t\right) dt}{\sqrt{\int {{q_{DNS}}\left(x,y,z,t\right)}^2 dt} \sqrt{\int {{q_{est}}\left(x,y,z,t\right)}^2 dt}} \mbox{,}
	\label{eq:correlation}
\end{equation}
\noindent where $t$ is the time, $q$ denotes the perturbation component (e.g., pressure or a particular velocity or forcing component) and $y$ is the wall-normal coordinate.
Subscripts $est$ and $DNS$ denote estimate and simulation quantities, respectively.
To evaluate the correlation values, we employed probes traversing the channel wall-normal direction at $(x, z) = (0, 0)$ scanning the complete time series.
For all metrics computed in this work, the first and last $N_{fft}/2$ instants of the time series were discarded as they could not be estimated due to end effects when computing the convolution (\ref{eq:fq_time}) using finite data.

Both state- and forcing-component correlations are shown in figure \ref{fig:qf_correlation_Retau550} and the estimates were evaluated using true (continuous lines) and white-noise (dashed lines) forcing.
For the eddy-viscosity model (dash-dotted lines), only the streamwise velocity component is plotted, as similar behaviour is observed for the other state and forcing components.
Up to $y^{+} \approx 15$, the normalised correlation is higher than 0.95 and 0.65 for all state and forcing components, respectively, when using true forcing statistics.
For higher wall-normal distances the correlation decays, but values higher than 0.6 for the state components are obtained up to $y^{+} \approx 550$.
It is clear from the normalised correlation values that modelling the forcing as white noise is only accurate for distances very close to the wall, where the measurements are obtained.
For $y^{+} \geq 10$, the correlation drastically decays, reaching values below zero (the normalised correlations is defined between -1 and 1) for the streamwise velocity component at $y^{+} \approx 70$.
When the eddy-viscosity model is considered, lower correlation values than those observed for the white noise forcing estimates are observed near the wall, however, for distances higher than $y^{+} \approx 30$, the linear operator containing the eddy-viscosity displayed an intermediate correlation, between white and true forcing statistics estimates.
Forcing correlation values are not as good as those observed for the state components, reaching a maximum of approximately 0.85 at $y^{+} \approx 6$, for the streamwise forcing component, and stabilising at approximately 0.45 for $y^{+} \geq 100$ when considering true forcing statistics.

Note that the white-noise assumption produced a saw-tooth like curve for the estimated pressure.
Such a feature is also observed for the estimated pressure CSD (in frequency domain) and for the subsequent statistics presented in this paper and we conjecture that it is related to the white-noise forcing assumption.
Estimates with a quasi-white forcing CSD, with coherence length equal to twice the maximal grid spacing, eliminates such oscillations, but such an assumption is not employed in this study.

\begin{figure}
	\centerline{\includegraphics[width=\textwidth]{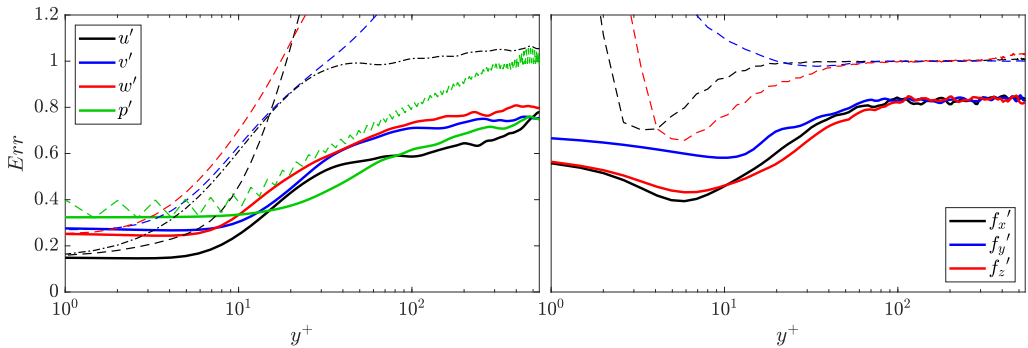}}
	\caption{Normalised RMS error between filtered DNS and estimate data for $Re_{\tau} \approx 550$ case and wall measurements of shear stress and pressure. See comments in the caption of figure \ref{fig:qf_correlation_Retau550}.}
	\label{fig:qf_RMS_error_Retau550}
\end{figure}

Figure \ref{fig:qf_RMS_error_Retau550} shows normalised RMS (root mean square) errors for state- and forcing-component estimates with respect to the filtered DNS.
The RMS error is given by 
\begin{equation}
	Err(x,y,z) = \frac{\sqrt{\int \left({q_{est}}\left(x,y,z,t\right) - {q_{DNS}}\left(x,y,z,t\right)\right)^2 dt}}{{\sqrt{\int {q_{DNS}}\left(x,y,z,t\right)^2 dt}}} \mbox{.}
	\label{eq:RMSerror}
\end{equation}

Lower errors are obtained from estimates using true forcing statistics (continuous lines) than white noise forcing (dashed lines).
Once again, only the streamwise velocity component is shown for the eddy-viscosity model (dashed lines), which reduced the error compared to the white noise forcing estimate.
These results are in accordance with the previous correlation values (figure \ref{fig:qf_correlation_Retau550}) and estimated snapshots (figure \ref{fig:q_estimation_Retau550_yplus15} through \ref{fig:q_estimation_Retau550_yplus200} and \ref{fig:u_nuT_Retau550}).
For the state estimates obtained using true forcing statistics, near the wall the normalised error is approximately constant up to $y^{+} \approx 5$ and then increases towards the channel center.
Interestingly, the normalised error at $y^{+} \approx 550$ is approximately 0.77 for all state components.
For the three forcing components, the normalised error initially decays, up to $y^{+} \approx 10$, then reaches a maximum value of approximately 0.82 at $y^{+} \approx 80$ which remains almost constant up to $y^{+} \approx 550$.
White-noise forcing and the eddy-viscosity model assumptions produce high errors that far exceed the axis limits selected to highlight the results obtained using true forcing statistics and the eddy-viscosity model (peak error values for white noise forcing were 46.12, 1.60, 2.82, 790.64, 20789 and 1590 for $u^{\prime}$, $v^{\prime}$, $w^{\prime}$, ${f_x}^{\prime}$, ${f_y}^{\prime}$, and ${f_z}^{\prime}$, respectively).

\begin{figure}
	\centerline{\includegraphics[width=\textwidth]{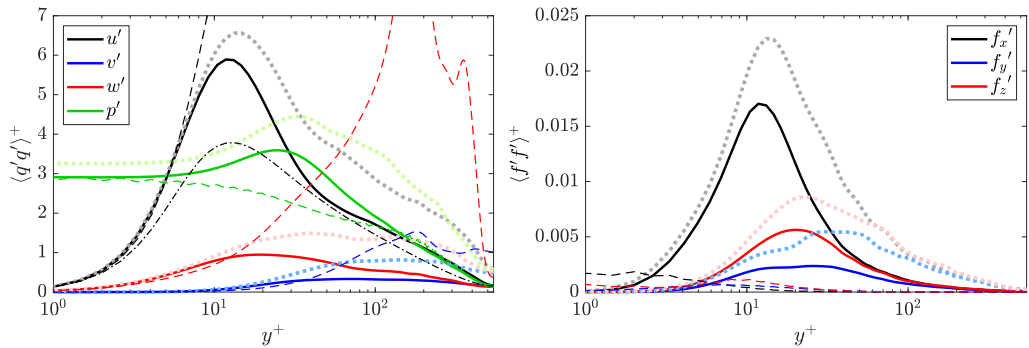}}
	\caption{Variance of filtered DNS and estimate data for $Re_{\tau} \approx 550$ and wall measurements of shear stress and pressure. See comments in the caption of figure \ref{fig:qf_correlation_Retau550}. The thick light dotted lines correspond to DNS values. Continuous, dashed and dash-dotted lines denote estimates obtained with true forcing statistics, white-noise and the eddy-viscosity model, respectively.}
	\label{fig:qf_variance_Retau550}
\end{figure}

The variances of the state and forcing components are displayed in figure \ref{fig:qf_variance_Retau550}.
Filtered DNS values (thick light dotted lines), true (continuous lines), white noise (dashed lines) and eddy-viscosity model (dash-dotted lines) estimates are provided in the figure.
As the filtered DNS values are used, the variances are lower than those of the full DNS (reported in \citet{morra2021colour}), but they are useful for comparison with the estimates, as both include the same wavenumbers.
The variance of the estimates obtained using the true forcing statistics are close to the DNS values, with slight underestimations for $u^{\prime}$ and $p^{\prime}$ and more significant underestimations for $v^{\prime}$ and $w^{\prime}$, whereas comparisons with the white-noise forcing estimates are not as good.
The eddy-viscosity model, for which only the streamwise velocity component results are shown, corrected the overshoot of white-noise forcing estimate but underestimates the variance values.
Again, the peak error values for white-noise forcing far exceed the axes (2846 and 9.55 for $u^{\prime}$ and $w^{\prime}$, respectively).
Two main factor contribute to the underestimation of the variance when employing the true statistics to build the transfer function.
First, if a structure has no footprints on the wall, it cannot be estimated.
Second, if there are more uncorrelated structures that contribute to the variance than there are sensors (2-3 in our case), not all of them will be properly estimated.

\begin{figure}
	\centerline{\includegraphics[width=\textwidth]{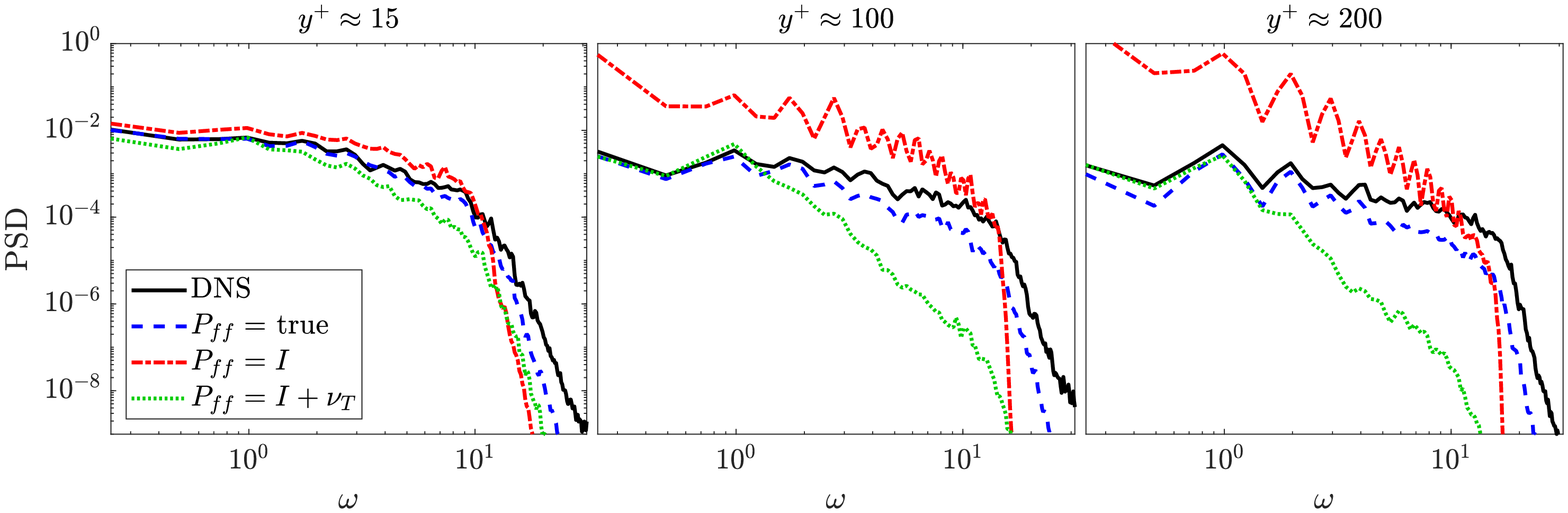}}
	\caption{Streamwise velocity PSD for $Re_{\tau} \approx 550$ and wall measurements of shear stress and pressure. Measurements taken for the probe located at $(x, z) = (0, 0)$ as a function of wall-normal direction $y^{+}$. Frames, from left to right: $y^{+} \approx 15$, 100 and 200. The frequency $\omega$ is non-dimensionalised in outer units.}
	\label{fig:q_PSD_Retau550}
\end{figure}

Figure \ref{fig:q_PSD_Retau550} shows the PSD of the streamwise velocity as a function of non-dimensional frequency $\omega$ evaluated for $y^{+} \approx 15$, 100 and 200.
DNS results (continuous lines) and true (dashed lines), white (dash-dotted lines) and eddy-viscosity model (dotted lines) forcing estimates are displayed in the figure.
Near the wall, the PSD of the estimates obtained using true forcing statistics is close to the DNS results, especially up to $\omega \approx 10$.
The white-noise forcing assumption slightly overestimates the PSD at lower frequencies and for higher frequencies, i.e., $\omega \geq 12$, the spectral content is underestimated.
When considering the white-noise estimate, the low- and high-frequency PSD levels are considerably overestimated and underestimated, respectively, for larger wall-normal distances.
The eddy-viscosity model estimate achieves an intermediate performance, closer to the true-forcing-statistics assumption for lower frequencies and to the white-noise assumption for higher frequencies.
This analysis corroborates the previous results of this section and shows that the resolvent-based estimator is best suited for low frequency flow dynamics.
Results for other state components are qualitatively similar.

\subsection{Comparison between modelling strategies for the forcing}
\label{sec:Retau550_forcing modelling}

In the above sections, three different modelling strategies for the unknown forcing terms were studied, i.e., the true forcing statistics, obtained directly from the DNS, white noise forcing and coloured noise based on an eddy-viscosity model.
Table \ref{tab:forcing_modelling_efforts} summarises the main features of each of these strategies, focusing on the computational cost and estimation accuracy.

\begin{table}
  \begin{center}
		\def~{\hphantom{0}}
		\begin{tabular}{c c c c}
			\textbf{Modelling strategy}	& \textbf{Computational cost}		& \makecell{\textbf{Estimate accuracy}\\\textbf{near-wall structures}}	& \makecell{\textbf{Estimate accuracy}\\\textbf{large structures}}	\\ [3pt]
			\hline
			true forcing								& high													& high																																	& mid-high	\\ [3pt]
			white noise									& low														& mid-high																															&	low				\\ [3pt]
			eddy-viscosity							& low														& middle																																&	mid-low		\\ [3pt]
			\hline
		\end{tabular}
		\caption{Summary of the cost and accuracy of the three forcing terms modelling strategies.}
		\label{tab:forcing_modelling_efforts}
  \end{center}
\end{table}

As the problem studied here has two homogeneous directions, estimates were obtained individually for each wavenumber pair, wall-normal discretisation, and time step.
The larger significant cost for this problem is constructing $\mathsfbi{P_{ff}}$ from the DNS data using Welch's method.
With this information, constructing the estimator (\ref{eq:ifft_TfTq_time}) is very inexpensive, requiring only matrix multiplications of size $4 N_y$, and the convolutions (\ref{eq:fq_time})-(\ref{eq:ifft_ab}) can be effectively computed using fast Fourier transforms.
For an effective implementation of this procedure for large databases, we refer the reader to \citet{martini2020resolvent}.

As shown in the previous subsections, the true forcing statistics provide greater accuracy compared to the other two forcing models.
For near-wall structures, the estimate is very similar to the DNS when the true forcing is considered, closely followed by white-noise forcing.
Interestingly, for near-wall structures, the white-noise forcing provided more accurate estimates than the eddy-viscosity model.
On the other hand, estimates of large-scale structures are not as accurate, although the coloured forcing outperformed the other two models.
The eddy-viscosity model provided better estimates than white noise forcing for larger scale structures.
The performance is nonetheless considerably worse than when true forcing statistics are used.

\section{Effect of Reynolds number}
\label{sec:Retau180_1000}

To assess the impact of the Reynolds number, estimations of channel flow at $Re_{\tau} \approx 180$ and 1000 were also performed.
For the $Re_{\tau} \approx 1000$ case, only wall measurements of skin friction were available to build the transfer functions; however, figure \ref{fig:u_C_Retau550} shows that the estimates remain accurate when only the shear components are considered.
As previously addressed in \S \ref{sec:Retau550_sensors}, there is a gain in the estimator performance by adding the wall pressure component to the observations, but it is better to retain the wall shear stress alone than the wall pressure.
Hence, to enable a fair comparison among the three simulations, i.e., $Re_{\tau} \approx 180$, 550 and 1000, the estimates shown in this section were obtained considering only skin friction on the channel walls.
Regarding the choice of forcing terms, true forcing statistics (obtained from the DNS) were employed for all results presented in this section, as the inclusion of accurate forcing colour proved to be crucial for reliable estimates far from the wall.
Appendix \ref{app:Retau180_1000} contains results for $Re_{\tau} \approx 180$ and 1000 analogous to those presented in \S \ref{sec:Retau550} for $Re_{\tau} \approx 550$.
In what follows we present metrics that allow a direct comparison of estimation accuracy for the three Reynolds numbers, illustrating the effect of $Re_{\tau}$ on the results.

\subsection{Overall comparison of estimation performance}
\label{sec:Retau180_1000_performance}

\begin{figure}
	\centerline{\includegraphics[width=\textwidth]{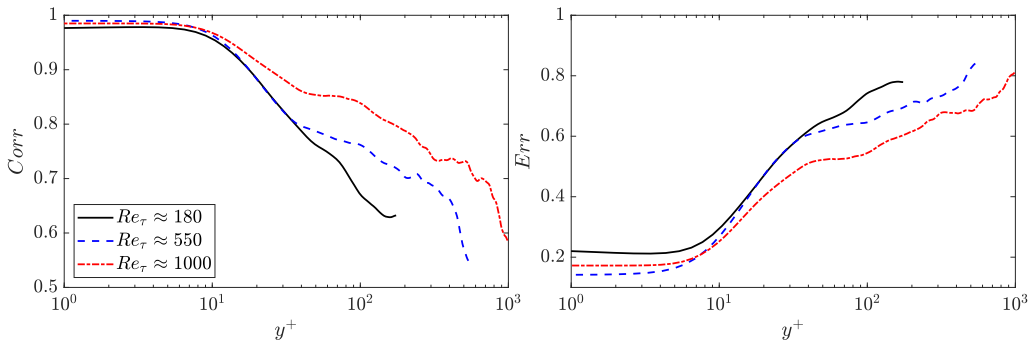}}
	\caption{Streamwise velocity correlation (left frame) and error (right frame) evaluated for the probe at $(x,~z) = (0,~0)$. Estimates obtained using wall measurements of shear stress and true forcing statistics for $Re_{\tau} \approx$ 180, 550 and 1000.}
	\label{fig:q_corr_err_Retau180_550_1000}
\end{figure}

Figure \ref{fig:q_corr_err_Retau180_550_1000} shows results of correlation ($Corr$, defined in \ref{eq:correlation}) and normalised RMS error ($Err$, defined in \ref{eq:RMSerror}) for the streamwise velocity component for the three Reynolds numbers, an estimates obtained using true forcing statistics.
In the figure, the continuous, dashed and dash-dotted lines denote the $Re_{\tau} \approx 180$, 550 and 1000 cases, respectively.
Very close to the wall, the $Re_{\tau} \approx 550$ estimate achieves the highest correlation value, of approximately 0.99, followed by the $Re_{\tau} \approx 1000$ and 180 estimate, which have maximum correlation values of 0.98 and 0.97, respectively.
The high correlations are almost constant up to $y^{+} \approx 7$ for all three Reynolds numbers.
For higher distances, the correlation deteriorates for all cases.
At the channel centre, the correlation decreases to approximately 0.6 for the $Re_{\tau} \approx 180$, 550 and 1000 estimates, respectively.
Despite the decrease in correlation between estimates and DNS data, the values remain significant throughout the channel for all three Reynolds numbers.
A collapse of correlations until the end of the buffer layer, at $y^{+} \approx 30$, is obtained for $Re_{\tau} \approx 180$ and 550 estimates.
Similar trends as those observed in figure \ref{fig:q_corr_err_Retau180_550_1000} were obtained for the other state components ($v$, $w$ and $p$).

Similar properties of the estimates can be inferred using the normalised RMS error.
Close to the wall, the lowest errors are observed for $Re_{tau} \approx 550$ (approximately 0.14), followed by $Re_{\tau}\approx 1000$ and 180 (0.17 and 0.22, respectively).
The hierarchy of estimation performance for the RMS normalised error metric is the same as that observed for the correlation metric, as well as the collapse of $Re_{\tau} \approx 180$ and 550 cases and the slope for the three cases in the $15 \leq y^+ \leq 35$ range.
Moving towards the channel center, the error metric deteriorates and is approximately 0.8 for the $Re_{\tau} \approx 180$, 550 and 1000 estimates, respectively.
Again, similar trends for the other state components were obtained.

The overall correlation and error shown in figure \ref{fig:q_corr_err_Retau180_550_1000} helps establish a Reynolds number trend, but there are differences in the wavenumbers retained in each database (presented in table \ref{tab:cutoff_parameters}), such that the range of wavenumbers may lead to differences in estimate accuracy.
Moreover, the wavenumber spectra is known to display an increasing outer peak, related to large-scale structures, as the Reynolds number is increased \citep{smits2011high}.
In what follows we will assess the estimation accuracy for individual wavenumbers to better evaluate the Reynolds-number trend.

\subsection{Estimation accuracy as a function of wavenumber}
\label{sec:Retau180_1000_wavenumber}

Normalized correlation and RMS error metrics as a function of each wavenumber pair ($\alpha,~\beta$) and channel height $y$ can be obtained as
\begin{equation}
	Corr(\alpha,y,\beta) = \frac{\int {q_{DNS}}\left(\alpha,y,\beta,t\right) {q_{est}^{*}}\left(\alpha,y,\beta,t\right) dt}{\sqrt{\int {{q_{DNS}}\left(\alpha,y,\beta,t\right)}^2 dt} \sqrt{\int {{q_{est}}\left(\alpha,y,\beta,t\right)}^2 dt}} \mbox{,}
	\label{eq:Corr_alpha_beta}
\end{equation}
\noindent and
\begin{equation}
	Err(\alpha,y,\beta) = \frac{\sqrt{\int \sum_{i=1}^{3} \left|{q_{est}^i}\left(\alpha,y,\beta,t\right) - {q_{DNS}^i}\left(\alpha,y,\beta,t\right)\right|^2 dt}}{{\sqrt{\int \sum_{i=1}^{3} \left|{q_{DNS}^i}\left(\alpha,y,\beta,t\right)\right|^2 dt}}} \mbox{.}
	\label{eq:RMSerror_alpha_beta}
\end{equation}
The superscript $i$ in ${q^{i}}$, for $i = 1$, 2 and 3, denotes streamwise ($u^{\prime}$), wall-normal ($v^{\prime}$) and spanwise ($w^{\prime}$) velocity components, respectively.
Note that instead of employing the probe at $(x,~z) = (0,~0)$, as in (\ref{eq:correlation}) and (\ref{eq:RMSerror}), the error and correlation curves obtained after (\ref{eq:RMSerror_alpha_beta}) and (\ref{eq:Corr_alpha_beta}) are functions of each wavenumber pair and wall-normal distance $y$.
This error metric is the same as that employed by \citet{illingworth2018estimating} and \citet{oehler2018linear} to evaluate the performance of their estimator for each wavenumber.

\begin{figure}
	\centerline{\includegraphics[width=\textwidth]{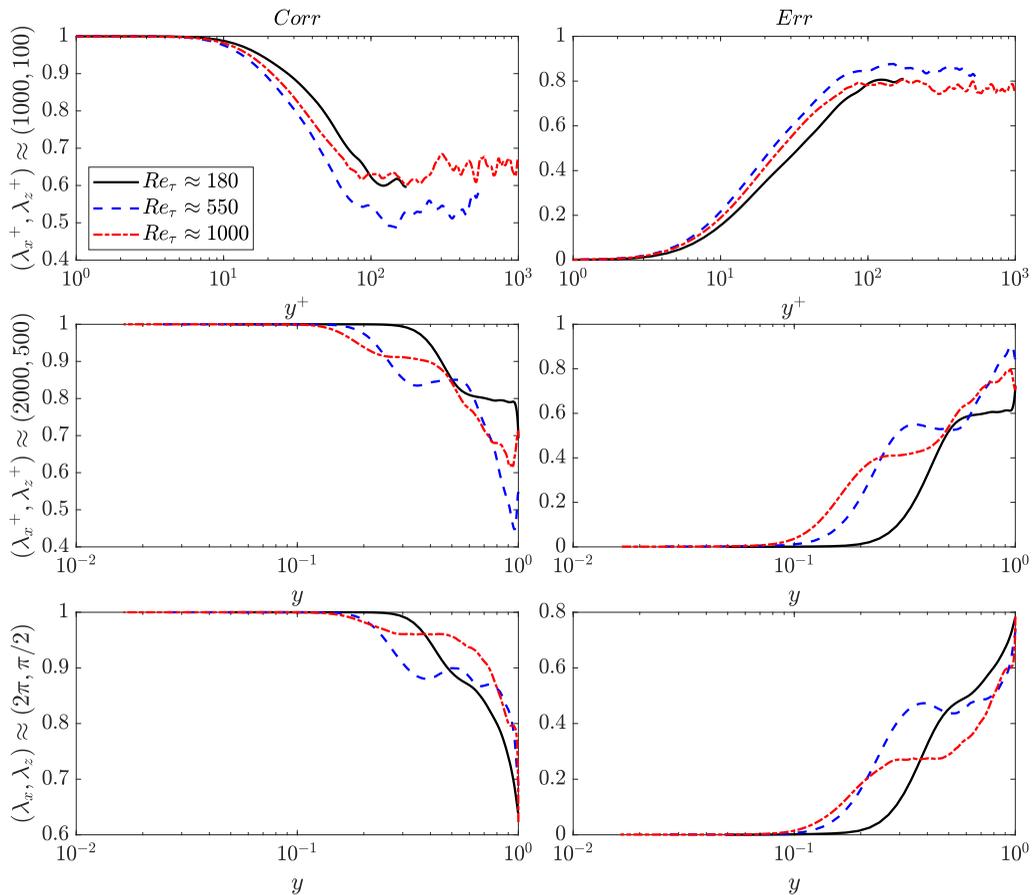}}
	\caption{Streamwise velocity correlation (\ref{eq:Corr_alpha_beta}, left frame) and error (\ref{eq:RMSerror_alpha_beta}, right frame) evaluated for $({\lambda_{x}}^{+},~{\lambda_{z}}^{+}) \approx (1000,~100)$ (top row), $({\lambda_{x}}^{+},~{\lambda_{z}}^{+}) \approx (2000,~500)$ (middle row) and $({\lambda_{x}},~{\lambda_{z}}) \approx (2\pi,~\pi/2)$ (bottom row). Estimates obtained using wall measurements of shear stress and true forcing statistics for $Re_{\tau} \approx$ 180, 550 and 1000.}
	\label{fig:q_corr_err_alpha_beta_Retau180_550_1000}
\end{figure}

Figure \ref{fig:q_corr_err_alpha_beta_Retau180_550_1000} shows correlation and normalised RMS error for $({\lambda_{x}}^{+},~{\lambda_{z}}^{+}) \approx (1000,~100)$ (top row), representative of near-wall structures, $({\lambda_{x}},~{\lambda_{z}}) \approx (2\pi,~\pi/2)$ (bottom row), representative of large-scale structures, and $({\lambda_{x}}^{+},~{\lambda_{z}}^{+}) \approx (2000,~500)$ (middle row), representative of intermediate-size structures, typical of a log layer if the Reynolds number is sufficiently high.
For near-wall structures, correlation and error are approximately independent of the Reynolds numbers (especially the error) if wall-normal distance is scaled in inner units.
On the other hand, the correlation and error results for the three Reynolds numbers do not collapse as well for the large-scale structures, but similar trends are observed if outer units are used to scale the wall-normal coordinate.
This is in agreement with the expected scaling of buffer layer and large-scale structures in inner and outer units, respectively \citep{smits2011high}.

\begin{figure}
	\centerline{\includegraphics[width=\textwidth]{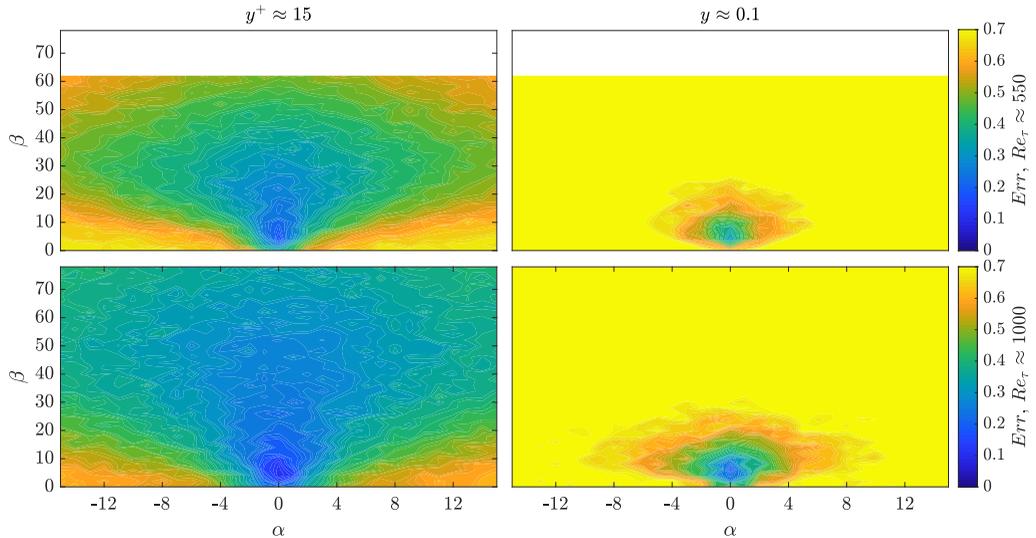}}
	\caption{Normalised RMS error (\ref{eq:RMSerror_alpha_beta}) as a function of $(\alpha,~\beta)$ for $y^+ \approx 15$ (left column) and $y \approx 0.1$ (right column). Top and bottom rows show results for $Re_{\tau} \approx$ 550 and 1000. True forcing statistics and shear stresses on both walls were used to perform the estimates.}
	\label{fig:alpha_beta_error_Retau550_1000}
\end{figure}

Figure \ref{fig:alpha_beta_error_Retau550_1000} shows the normalised RMS error, evaluated according to (\ref{eq:RMSerror_alpha_beta}), for planes $y^+ \approx 15$ and $y \approx 0.1$, left and right frames, respectively, and for the $Re_{\tau} \approx$ 550 and 1000 estimates, top and bottom frames, respectively.
In the plots, the contour scale is limited between 0 and 0.7, to enable a comparison with the plots by \citet{illingworth2018estimating} and \citet{oehler2018linear}.
For the $Re_{\tau} \approx$ 550 estimates, the maximum errors are approximately 0.68 and 0.86 for planes $y^+ \approx 15$ and $y \approx 0.1$, respectively, whereas for the $Re_{\tau} \approx$ 1000 estimates, the maximum errors are approximately 0.62 and 0.79 for planes $y^+ \approx 15$ and $y \approx 0.1$, respectively.
The largest errors occur for $\beta \approx 0$ and high values of $\alpha$, i.e., structures that are long in the spanwise direction and thin in the streamwise direction.
Elongated structures in the streamwise direction, with low $\alpha$, are more accurately estimated.

\begin{figure}
	\centerline{\includegraphics[width=\textwidth]{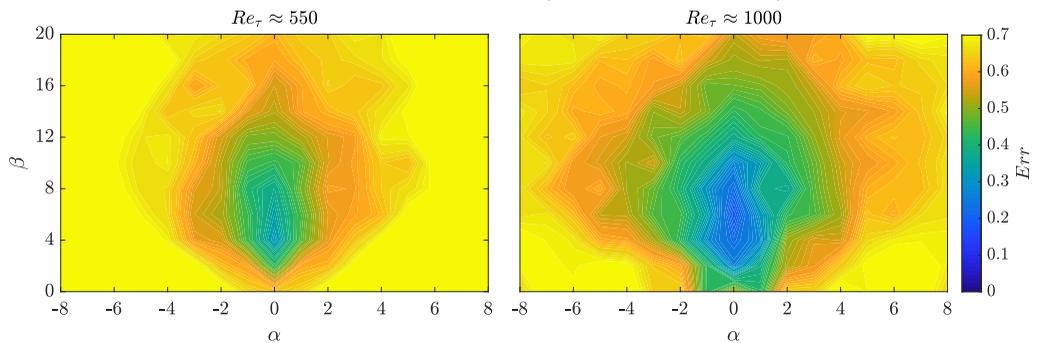}}
	\caption{Zoomed in version of figure \ref{fig:alpha_beta_error_Retau550_1000}, with the normalised error for $y \approx 0.1$, restricting the wavenumber space to $-8 \leq \alpha \leq 8$ and $0 \leq \beta \leq 20$.}
	\label{fig:alpha_beta_error_Retau550_1000_zoom}
\end{figure}

\citet{illingworth2018estimating} employed a Kalman filter and observed the streamwise, wall-normal and spanwise velocity components at a wall-normal distance of $y^+ = 197$ as inputs, evaluating the normalised RMS the error at the plane $y \approx 0.1$ ($y^+ = 100$ for the $Re_{\tau} = 1000$ channel considered in that work).
The authors studied two estimators, the first based on the linearised Navier-Stokes (LNS) equations with white-noise forcing, which showed a maximum error of 8.37, and the second including an eddy-viscosity model in the linear operator, which boosted the performance, displaying a maximum error of 1.07.
Figure \ref{fig:alpha_beta_error_Retau550_1000_zoom} contains a zoomed in version of the bottom frames of figure \ref{fig:alpha_beta_error_Retau550_1000}, restricting the wavenumber axes to match figure 4 of \citet{illingworth2018estimating}, i.e., $-8 \leq \alpha \leq 8$ and $0 \leq \beta \leq 20$.
The right frame of figure \ref{fig:alpha_beta_error_Retau550_1000} is equivalent to figures 4(a)-(b) in \citet{illingworth2018estimating}, i.e., both figures were obtained for $Re_{\tau} \approx$ 1000, with contour scale and wavenumber space limited to $Err \in [0,~0.7]$, $\alpha \in [-8,~8]$ and $\beta \in [0,~20]$.
The present estimates provide much lower errors than those obtained by \citet{illingworth2018estimating}, especially in the neighbourhood of $\alpha \approx 0$.

\begin{figure}
	\centerline{\includegraphics[width=0.49\textwidth]{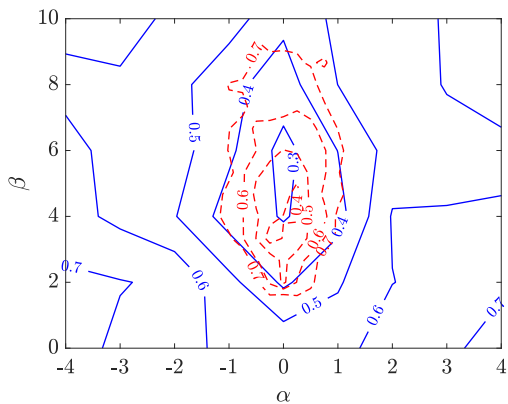}}
	\caption{Comparison between the normalised RMS errors of the $Re_{\tau} \approx 1000$ estimate (blue continuous contours) and \citet{oehler2018linear} results (red dashed contours), for estimates at $y = 0.1$ and 0.3. An average of $Err$ at the two heights was taken, following \citet{oehler2018linear}.}
	\label{fig:alpha_beta_error_Retau_1000_comparison}
\end{figure}

As discussed in the Introduction, the work by \citet{oehler2018linear} is an extension of \citet{illingworth2018estimating}, including estimates performed at $Re_{\tau} = 2000$.
The former authors also displayed normalised estimation error maps as a function of $(\alpha,~\beta)$ similar to those by \citet{illingworth2018estimating} in their figure 6.
Figure \ref{fig:alpha_beta_error_Retau_1000_comparison} shows a direct comparison between the normalized RMS error as a function of wavenumber pair for the $Re_{\tau} \approx 1000$ estimate (continuous contours) and the results by \citet{oehler2018linear} (dashed contours).
The resolvent-based method provides superior estimates, with lower error values extending over a wider wavenumber range.
The results in \S \ref{sec:Retau550} show that these lower errors are achieved here by incorporating forcing statistics taken from the DNS.
However, care should be taken in these comparisons with \citet{illingworth2018estimating} and \citet{oehler2018linear} due to differences in the measurement plane and Reynolds number, respectively.
Moreover, the procedure in \citet{illingworth2018estimating} and \citet{oehler2018linear} ensures a causal estimation, whereas the present estimates are not causal.

\subsection{Maximum observed height}
\label{sec:Retau180_1000_hObs}

\begin{figure}
	\centerline{\includegraphics[width=\textwidth]{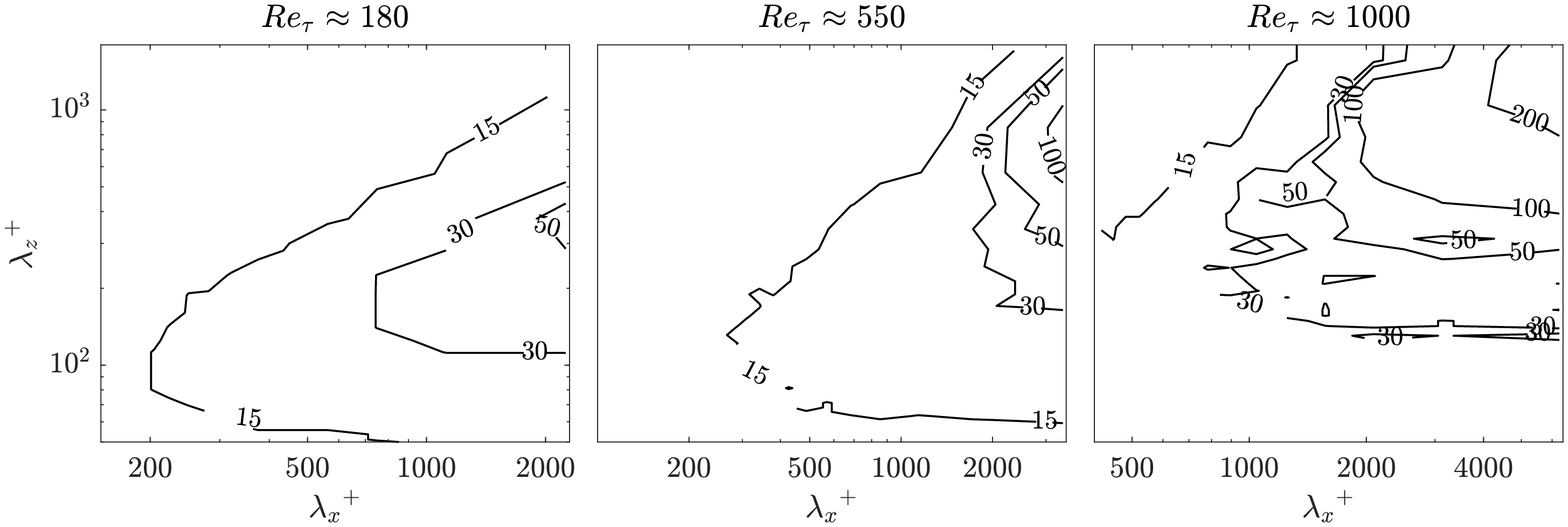}}
	\caption{Observed height ($y_{obs}^{+}$) maps for the three Reynolds numbers. Estimates obtained using wall measurements of shear stress for $Re_{\tau} \approx$ 180, 550 and 1000.}
	\label{fig:hObs_maps}
\end{figure}

The trend of accuracy loss far from the wall can be quantified by defining a maximum observed distance.
This distance is here defined as the position for which $Err(\alpha,{y_{obs}},\beta) \leq 0.5$.
Figure \ref{fig:hObs_maps} shows contour maps of the observable height in inner units, i.e, $y_{obs}^{+}$, measured from the channel wall, for the three studied Reynolds numbers.
This 0.5 maximum normalized error criterion and contour levels of $y^{+} = 15$, 30, 50, 100 and 200 were employed to obtain the maps shown in figure \ref{fig:hObs_maps}.
Smaller structures, of small wavelength pair (${\lambda_{x}}^{+},~{\lambda_{z}}^{+}$), can only be estimated near the wall.
Far from the wall, only the larger structures are observed.
This behaviour is consistently observed for the three Reynolds numbers.
These results corroborate the snapshot reconstructions and power spectra previously addressed, e.g., figure \ref{fig:u_C_Retau550}, where the small structures that are present in the DNS snapshot at a distance of $y^{+} \approx 200$ could not be properly estimated.
As the Reynolds number is increased, the emergence of larger-scale structures at higher $y^+$ is captured by the estimation, with large wavelengths observed at log-layer heights $y^+ \approx 100$ and 200.

\section{Conclusions}
\label{sec:conclusions}

A resolvent-based methodology was applied to obtain estimates of turbulent channel flow fluctuations at friction Reynolds numbers of approximately 180, 550 and 1000, using DNS data.
Wall measurements of skin friction (shear stress) and pressure were employed as observable quantities.
\citet{towne2020resolvent} and \citet{martini2020resolvent} developed the resolvent-based estimator and applied it to benchmark cases, demonstrating its feasibility to deal with channel flow problems, among others.
Here, the formulation was used to estimate flow fluctuations for the three Reynolds numbers at various wall-normal positions, and a thorough analysis of the performance of each estimation at different locations, investigating its frequency and wavenumber dependency, is provided.
The approach is an optimal linear estimation, as shown in \citet{martini2020resolvent}, if constructed with the true statistics from non-linear terms, which act as a forcing in the resolvent framework. 
Suboptimal estimators, which were obtained using a spatial-temporal white noise forcing with and without an eddy-viscosity model, were also studied. 

Accurate estimates were obtained in the near-wall region, up to $y^{+} \approx 15$, with correlations higher than 0.95 for $Re_{\tau} \approx 180$ and 550 and higher than 0.85 for $Re_{\tau} \approx 1000$ when using the true forcing statistics computed from DNS data.
Most of the structures present in the DNS snapshots were recovered by the estimates at these wall-normal positions.
Even for transfer functions based on white-noise forcing statistics, good estimates were obtained in the buffer layer.
Comparisons of power spectra of DNS and estimates showed good agreement, particularly in the buffer layer.
The peak in the power spectra of the streamwise velocity, located at $({\lambda_x}^{+},~{\lambda_z}^{+}) \approx (1000,~100)$, was well recovered for all friction Reynolds numbers studied.
Estimates were not computed for wavelengths lower than $({\lambda_x}^{+},~{\lambda_z}^{+}) \approx (1000,~100)$.
Power spectra of the other flow variables were also well estimated.

The estimation quality deteriorates far from the wall, although many structures were still present in the estimated snapshots which are solely based on wall measurements.
It can be inferred that many larger-scale structures present in the outer layer leave their footprint on the wall, as reported in earlier works (see for instance \citet{smits2011high}, \citet{jimenez2013near} and references therein).
The resolvent-based estimator is able to reconstruct such large-scale structures with reasonable accuracy when true forcing statistics, computed from DNS data, are employed to build the transfer function.
Corroborating previous studies \citep{morra2021colour, martini2020resolvent}, the use of spatially white noise forcing did not provide good results for higher wall-normal positions.
The agreement between DNS and estimated power spectra deteriorates far from the wall.

The present results shed some light on the need to include an eddy-viscosity model in the linearised operator to build the estimators of \citet{illingworth2018estimating} and \citet{towne2020resolvent}. 
The results of this work show poor performances at distances far from the wall if white-noise forcing is assumed without eddy-viscosity.
\citet{morra2019relevance} and  \citet{morra2021colour}, respectively, showed that including an eddy-viscosity leads to improved predictions by the resolvent and that this improvement can be attributed to an implicit colouring of the effective forcing by the eddy-viscosity model.
Hence, the results by \citet{illingworth2018estimating} and \citet{towne2020resolvent} may be seen as an improvement over modelling non-linear terms when more detailed information on these are unknown, as part of the forcing colour is incorporated in the linear operator by the eddy-viscosity model.
This effect was seen in the present work, as a white-noise-forced resolvent estimation including eddy-viscosity has intermediate accuracy, improving predictions from the white-noise model with molecular viscosity but with poorer performance than the same model including the forcing statistics from the DNS.
Such forcing colour is here seen to be crucial to obtain accurate estimates of flow fluctuations, especially velocity components further from the wall.
If true forcing statistics are unknown, a low-rank forcing model can be constructed using an auxiliary set of sensors \citep{martini2020resolvent}.
If such forcing is unavailable, it should be modelled, and the present results show that, while the use of an eddy-viscosity model improves estimates far from the wall if compared to white-noise forcing, it worsens predictions close to the wall.

Overall, the present study adds one more tool for estimating the flow state from noisy low-rank measurements.
Such a tool has potential to be employed for flow-control problems, where the controller has to have information on the flow state at each time step in order to apply a control law that, for example, reduces the skin friction or turbulent kinetic energy.
Although the present estimations are not causal, they may be modified to ensure causality by a Wiener-Hopf procedure \citep{martinelli2009feedback, martini2019linear, jung2020optimal}.
Further work in that direction is ongoing.
The robustness of the resolvent-based methodology should also be evaluated in future work, e.g., by assessing the ability of a transfer function derived for one Reynolds number to estimate the flow at a different Reynolds number.
The observed robustness to changes in Reynolds number of the data-based transfer functions used by \citet{sasaki2019transfer} encourage us towards this direction.
Finally, the methodology naturally accommodates a global resolvent formulation instead of the parallel-flow formulation used in this study.
For channel flow, the base flow only depends on the wall-normal coordinate $y$, and all operations can be performed by forming matrices that discretise the problem.
More complex base flows may be considered using the procedures described in \citet{martini2020resolvent}, which lead to the required transfer functions using time steppers without forming or inverting matrices.


\noindent{\bf  Supplementary data\bf{.}} \label{SupMat} Supplementary material and movies are available at \\https://doi.org/10.1017/jfm.2020... \\

\noindent{\bf Acknowledgements\bf{.}} The authors thank Prof. Dan S. Henningson for the many fruitful contributions. We also thank Dr. Simon Illingworth and Dr. Stephan Oehler for sharing the data used in figure \ref{fig:alpha_beta_error_Retau_1000_comparison}. \\

\noindent{\bf Funding\bf{.}} F. R. Amaral received funding from from São Paulo Research Foundation (FAPESP/Brazil), grant \#2019/02203-2. A. V. G. Cavalieri was supported by the National Council for Scientific and Technological Development (CNPq/Brazil), grant \#310523/2017-6. A.T. was supported by the Air Force Office of Scientific Research (AFOSR/USA) grant \#FA9550-20-1-0214.\\

\noindent{\bf  Author ORCID\bf{.}} F. R. Amaral, https://orcid.org/0000-0003-1158-3216; A. V. G. Cavalieri, https://orcid.org/0000-0003-4283-0232; E. Martini, https://orcid.org/0000-0002-3144-5702; P. Jordan, https://orcid.org/0000-0001-8576-5587; A. Towne, https://orcid.org/0000-0002-7315-5375.\\


\noindent{\bf Declaration of Interests\bf{.}} The authors report no conflict of interest. \\


\appendix

\section{Resolvent operator}
\label{app:math}

In this appendix, we provide more details on the resolvent-based formulation.
The linearised Navier-Stokes (LNS) operator, denoted by $\mathsfbi{L} = (- i \omega \mathsfbi{M} - \mathsfbi{A})$ in (\ref{eq:state-space_time}) and (\ref{eq:state-space_freq}), is given by
\begin{equation}
	\mathsfbi{L} =
	\left[\begin{array}{cccc}
		i \alpha \mathsfbi{U} - i \omega \mathsfbi{I} - \frac{1}{Re} \boldsymbol{\nabla^2}	& \mathsfbi{\frac{d U}{d y}}	& \mathsfbi{0}	& i \alpha \mathsfbi{I} \\
		\mathsfbi{0} & i \alpha \mathsfbi{U} - i \omega \mathsfbi{I} - \frac{1}{Re} \boldsymbol{\nabla^2}	& \mathsfbi{0}	& \mathsfbi{\frac{d}{d y}} \\
		\mathsfbi{0} & \mathsfbi{0}	& i \alpha \mathsfbi{U} - i \omega \mathsfbi{I} - \frac{1}{Re} \boldsymbol{\nabla^2}	& i \beta \mathsfbi{I} \\
		i \alpha \mathsfbi{I}	& \mathsfbi{\frac{d}{d y}}	& i \beta \mathsfbi{I}	& \mathsfbi{0}
	\end{array}\right] \mbox{,}
	\label{eq:resolvent_operator}
\end{equation}
\noindent for each combination of $(\alpha,\beta,\omega)$.
In discretised form, the matrix has $4 N_y \times 4 N_y$ size, where $N_y$ is the number of points in the wall-normal direction.
In (\ref{eq:resolvent_operator}), $\mathsfbi{U}$ is the mean turbulent streamwise velocity profile (diagonal matrix of $N_y \times N_y$ size), $\mathsfbi{I}$ is the identity matrix (of $N_y \times N_y$ size), $k^2 = \alpha^2 + \beta^2$, $\boldsymbol{\nabla^2} = \left(\mathsfbi{\frac{d^2}{d y^2}} - k^2 \mathsfbi{I}\right)$ with $k^2 = \alpha^2 + \beta^2$, $\mathsfbi{\frac{d}{d y}}$ and $\mathsfbi{\frac{d^2}{d y^2}}$ are, respectively, the first and second differentiation matrices along the wall-normal coordinate (matrices of $N_y \times N_y$ size), and $\mathsfbi{0}$ is the zeros matrix (of $N_y \times N_y$ size).
No-slip conditions at the walls are applied to the linear operator.
Therefore, the resolvent operator, before the application of observation ($\mathsfbi{C}$) and actuation ($\mathsfbi{B}$) operators, is obtained as $\mathsfbi{R} = (- i \omega \mathsfbi{I} - \mathsfbi{A})^{-1} = \mathsfbi{L}^{-1}$.

\begin{figure}
	\centerline{\includegraphics[width=0.5\textwidth]{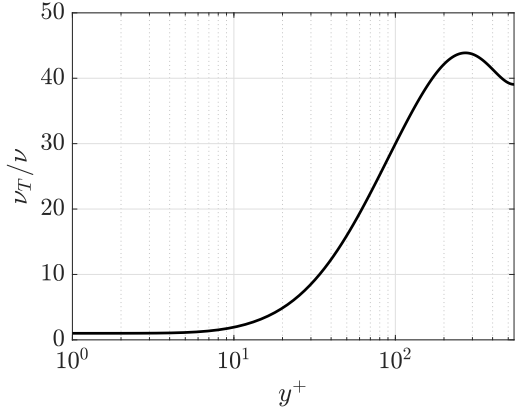}}
	\caption{Eddy-viscosity model for $Re_{\tau} \approx 550$.}
	\label{fig:EV_model_Retau550}
\end{figure}

When considering the eddy-viscosity model \citep{towne2020resolvent}, the linear operator $\mathsfbi{L}$ is given by 
\begin{equation}
	\mathsfbi{L} =
	\left[\begin{array}{cccc}
		i \alpha \mathsfbi{U} - i \omega \mathsfbi{I} - \mathsfbi{E} - \mathsfbi{Z}	& - i \alpha \nu_T^{\prime} \mathsfbi{I} + \mathsfbi{\frac{d U}{d y}}	& \mathsfbi{0}	& i \alpha \mathsfbi{I} \\
		\mathsfbi{0} & i \alpha \mathsfbi{U} - i \omega \mathsfbi{I} - 2 \mathsfbi{E} - \mathsfbi{Z}	& \mathsfbi{0}	& \mathsfbi{\frac{d}{d y}} \\
		\mathsfbi{0} & - i \beta \nu_T^{\prime} \mathsfbi{I}	& i \alpha \mathsfbi{U} - i \omega \mathsfbi{I} - \mathsfbi{E} - \mathsfbi{Z}	& i \beta \mathsfbi{I} \\
		i \alpha \mathsfbi{I}	& \mathsfbi{\frac{d}{d y}}	& i \beta \mathsfbi{I}	& \mathsfbi{0}
	\end{array}\right] \mbox{,}
	\label{eq:eddy-viscosity_operator}
\end{equation}
\noindent where $\mathsfbi{E} = \nu_T^{\prime} \mathsfbi{\frac{d}{d y}}$, $\mathsfbi{Z} = \frac{1}{Re_{\tau}} \frac{\nu_T}{\nu} \boldsymbol{\nabla^2}$, $\nu = \frac{H}{Re}$ is the kinematic viscosity and $H$ is the channel half-height.
The total eddy-viscosity $\nu_T(y)$, obtained after the sum of the turbulent eddy-viscosity $\nu_t(y)$ and the viscosity $\nu$, is modeled as
\begin{equation}
	\frac{\nu_T}{\nu} = \frac{1}{2} \left\{1 + \frac{\kappa^2 {Re_{\tau}}^2}{9} \left(1 - \eta^2\right)^2 \left(1 + 2 \eta^2\right)^2 \left[1 - e^{\left(|\eta| - 1\right) Re_{\tau} / A}\right]^2\right\}^{1/2} + \frac{1}{2} \mbox{,}
\end{equation}
\noindent where the constants $\kappa$ and $A$ are given as 0.426 and 25.4, respectively \citep{pujals2009note}, and $\eta$ is the non-dimensional wall distance in outer units.
Figure \ref{fig:EV_model_Retau550} shows the eddy-viscosity model for $Re_{\tau} \approx 550$ as a function of $y^{+}$ up to the half-channel height.

The observation matrix $\mathsfbi{C}$ has $N_s \times 4 N_y$ size, where $N_s$ is the number of sensors.
For example, considering channel wall skin frictions as observations leads to a $\mathsfbi{C}$ matrix composed of 4 rows ($\frac{du}{dy}\rvert_{wall}$ and $\frac{dw}{dy}\rvert_{wall}$ at both channel walls) by $4 N_y$ columns.
Including wall pressure ($p_{wall}$) at both channel walls to the observation adds two rows to matrix $\mathsfbi{C}$.

The input matrix $\mathsfbi{B}$ has $4 N_y \times 3 N_y$ size and is defined as
\begin{equation}
	\mathsfbi{B} =
	\left[\begin{array}{ccc}
		\mathsfbi{I}	& \mathsfbi{0}	& \mathsfbi{0} \\
		\mathsfbi{0}	& \mathsfbi{I}	& \mathsfbi{0} \\
		\mathsfbi{0}	& \mathsfbi{0}	& \mathsfbi{I} \\
		\mathsfbi{0}	& \mathsfbi{0}	& \mathsfbi{0}
	\end{array}\right] \mbox{.}
	\label{eq:actuation_matrix}
\end{equation}

To enforce the boundary conditions, we zero the lines in $\mathsfbi{B}$ corresponding to the wall positions for the three momentum equations.

The forcing terms $\boldsymbol{f} = [\boldsymbol{f_x}~\boldsymbol{f_y}~\boldsymbol{f_z}]^T$, obtained after the linearisation of the Navier-Stokes system, are
\begin{eqnarray}
	\boldsymbol{f_x} &=& - \boldsymbol{u^{\prime}}\frac{d \boldsymbol{u^{\prime}}}{d x} - \boldsymbol{v^{\prime}}\frac{d \boldsymbol{u^{\prime}}}{d y} - \boldsymbol{w^{\prime}}\frac{d \boldsymbol{u^{\prime}}}{d z} \mbox{,} \\
	\boldsymbol{f_y} &=& - \boldsymbol{u^{\prime}}\frac{d \boldsymbol{v^{\prime}}}{d x} - \boldsymbol{v^{\prime}}\frac{d \boldsymbol{v^{\prime}}}{d y} - \boldsymbol{w^{\prime}}\frac{d \boldsymbol{v^{\prime}}}{d z} \mbox{,} \\
	\boldsymbol{f_z} &=& - \boldsymbol{u^{\prime}}\frac{d \boldsymbol{w^{\prime}}}{d x} - \boldsymbol{v^{\prime}}\frac{d \boldsymbol{w^{\prime}}}{d y} - \boldsymbol{w^{\prime}}\frac{d \boldsymbol{w^{\prime}}}{d z} \mbox{.}
	\label{eq:forcing_terms}
\end{eqnarray}

\section{DNS filtering}
\label{app:DNS_filtering}

\begin{figure}
	\centerline{\includegraphics[width=0.7\textwidth]{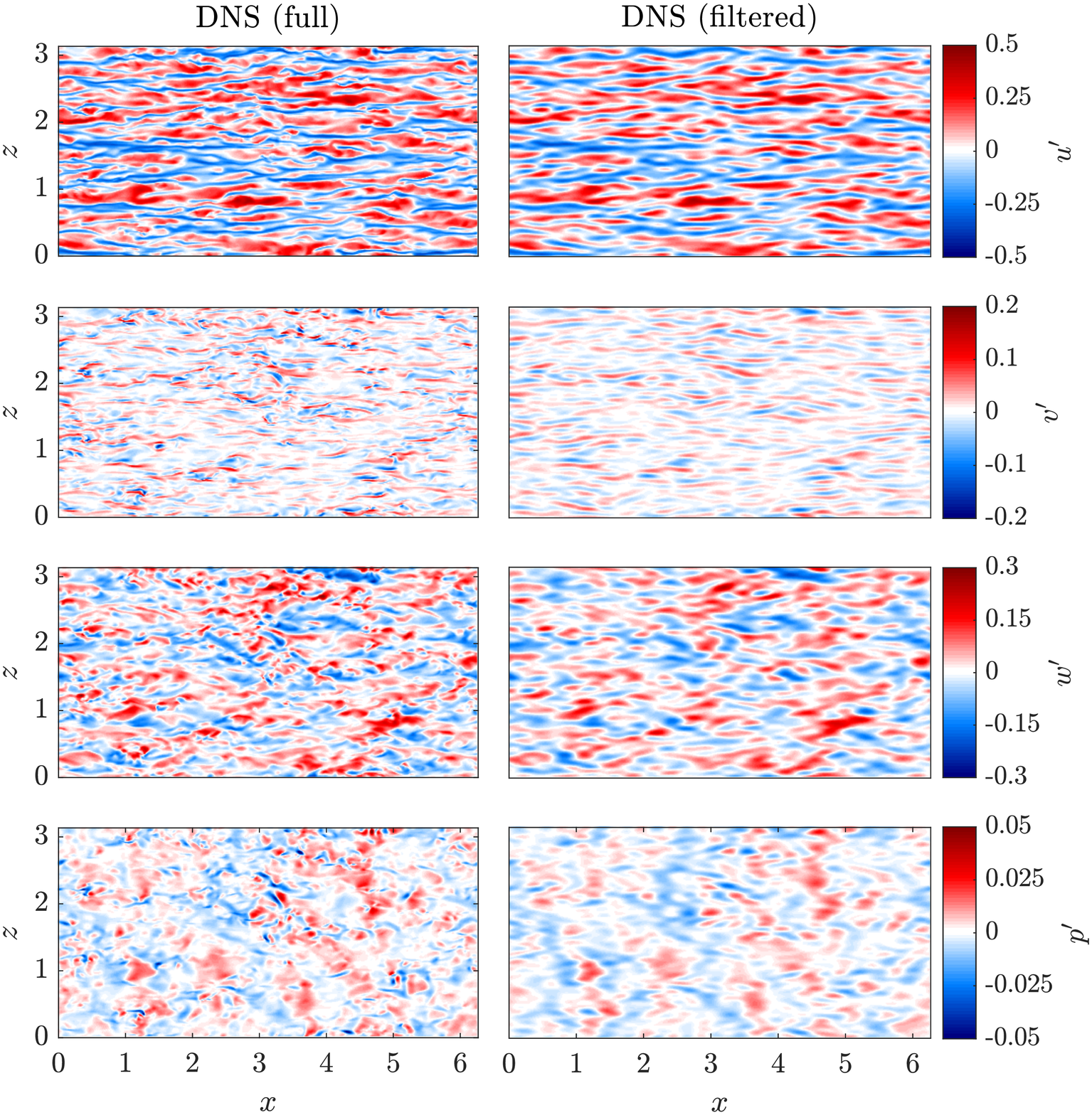}}
	\caption{Comparison between full and filtered DNS flow state snapshots for $Re_{\tau} \approx 550$, $y^{+} \approx 15$. Left column: full DNS data. Right column: filtered DNS data. Rows, from top to bottom: streamwise ($u^{\prime}$), wall-normal ($v^{\prime}$), spanwise ($w^{\prime}$) velocity fluctuations, respectively, and pressure fluctuation ($p^{\prime}$). Fluctuations shown in outer units.}
	\label{fig:q_DNS_Retau550_yplus15}
\end{figure}

\begin{figure}
	\centerline{\includegraphics[width=0.7\textwidth]{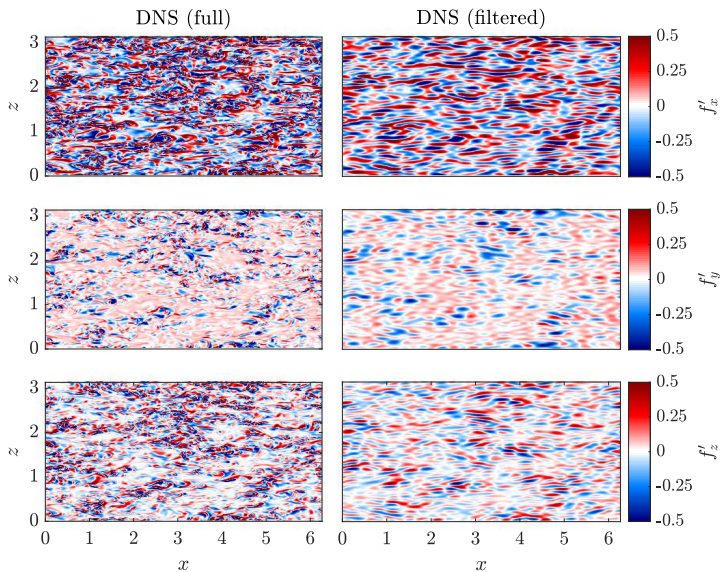}}
	\caption{Comparison between full and filtered DNS flow forcing snapshots for $Re_{\tau} \approx 550$, $y^{+} \approx 15$. Left column: full DNS data. Right column: filtered DNS data. Rows, from top to bottom: streamwise ($f_{x}^{\prime}$), wall-normal ($f_{y}^{\prime}$) and spanwise ($f_{z}^{\prime}$) forcing fluctuations, respectively. Fluctuations shown in outer units.}
	\label{fig:f_DNS_Retau550_yplus15}
\end{figure}

Figures \ref{fig:q_DNS_Retau550_yplus15} and \ref{fig:f_DNS_Retau550_yplus15} show a comparison between a full DNS snapshot, i.e., with no spatial wavenumber filtering, and the filtered DNS snapshot for $Re_{\tau} \approx 550$.
Results are provided for a wall-normal distance, based on wall units, of $y^{+} \approx 15$.
Flow state and forcing components are shown in the figures.
Most of the structures present in the flow were captured by the filtered DNS, although the smaller scale structures, of high wavenumber, were filtered out.
Qualitatively similar results were obtained for other snapshots, wall-normal distances and Reynolds numbers.
As addressed in \S \ref{sec:dns}, the filtered DNS data were used as input to the resolvent-based estimates through the wall measurements $\boldsymbol{y}$.

We defined a metric to evaluate the fraction of the total fluctuation energy retained after the filtering process as
\begin{equation}
	E(y) = \frac{\sum_1^{N_x \times Nz}\left({u_{filtered}^{\prime}}^2+{v_{filtered}^{\prime}}^2+{w_{filtered}^{\prime}}^2\right)}{\sum_1^{N_x \times Nz}\left({u_{full}^{\prime}}^2+{v_{full}^{\prime}}^2+{w_{full}^{\prime}}^2\right)} \mbox{,}
	\label{eq:retained_energy}
\end{equation}
\noindent where subscripts indicate the filtered and full DNS snapshots.
Note that the velocity fluctuation components at each grid point was squared and then summed up for a given wall distance $y$.

For the $Re_{\tau} \approx 550$ case, an energy fraction $E$ of approximately 0.86, 0.83 and 0.86 was obtained for $y^{+} \approx 15$, 90 and 200, respectively, after the filtering process.
On the other hand, the $Re_{\tau} \approx 180$ case displayed energy fraction values of approximately 0.91 and 0.85 at $y^{+} \approx 15$ and 90, whereas for the $Re_{\tau} \approx 1000$ case, the filtering process retained lower energy fractions, i.e., $E$ of approximately 0.70, 0.73 and 0.80 was obtained for $y^{+} \approx 15$, 100 and 200, respectively.

\section{Error metrics for different observation matrices}
\label{app:Cmatrix_errors}

\begin{figure}
	\centerline{\includegraphics[width=\textwidth]{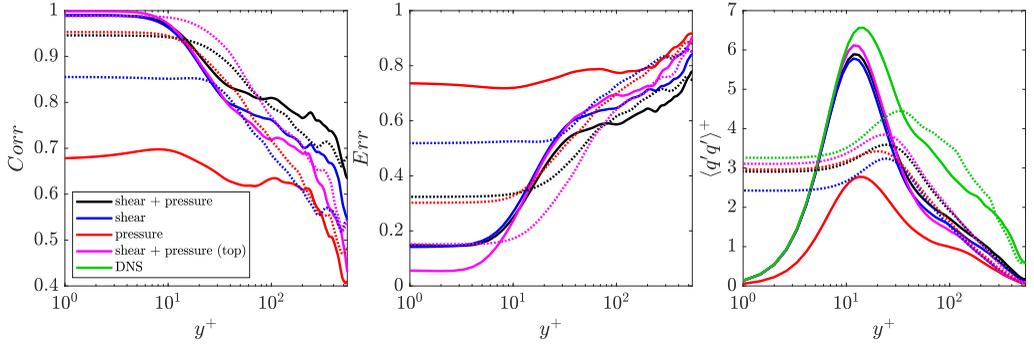}}
	\caption{Flow state comparison metrics for $Re_{\tau} \approx 550$ and different wall measurements for the streamwise velocity (continuous lines) and pressure (dotted lines). True forcing statistics were used to perform the estimates. Frames, from left to right: correlation, normalised RMS error and variance.}
	\label{fig:q_correlation_error_variance_Retau550}
\end{figure}

Figure \ref{fig:q_correlation_error_variance_Retau550} shows a comparison of error metrics for the estimates obtained using the true forcing statistics and different observation matrices.
Wall measurements of shear stress (blue lines), pressure (red lines), shear stress plus pressure at both walls (black lines) and at a single wall (magenta lines) are considered as observations.
Results for the streamwise velocity (continuous lines) and pressure (dotted lines) components are displayed in the figure.
The other velocity components, i.e., in the wall-normal and spanwise directions, displayed trends similar to the component in the streamwise direction.

The results corroborate the observations made after figure \ref{fig:u_C_Retau550}.
When both wall measurements are considered to perform the estimates, higher values of correlation, lower errors and variance closer to the DNS are obtained.
For the velocity components, estimations considering only the wall shear stress measurements are closer to the ones that take into account both wall quantities.
On the other hand, for the pressure component, estimates from observations of wall pressure outperform those from shear stress measurements.
When the sensors are placed at a single channel wall, near the wall containing the sensors, i.e., up to $y^+ \approx 20$, the estimates are more favourable, with higher correlation values, lower errors and variance closer to the DNS.
Far from the wall, the metrics deteriorate in comparison to those obtained with sensors on both walls.

\section{Additional results for $Re_{\tau} \approx$ 180 and 1000}
\label{app:Retau180_1000}

\begin{figure}
	\centerline{\includegraphics[width=\textwidth]{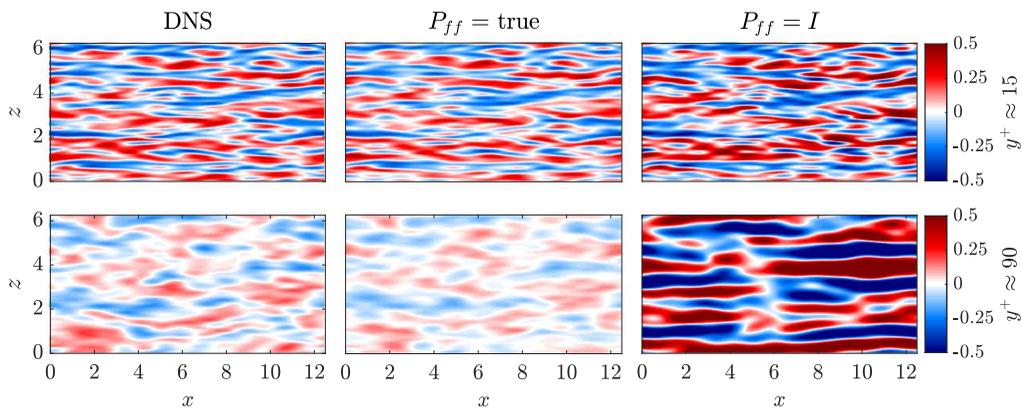}}
	\caption{Comparison between filtered DNS and resolvent-based estimates of a snapshot of the flow streamwise velocity for $Re_{\tau} \approx 180$ and wall measurements of shear stress and pressure. Top row: $y^{+} \approx 15$. Bottom row: $y^{+} \approx 90$. Columns, from left to right: DNS data, true and white noise forcing estimates. Fluctuations shown in outer units. An animated version of this figure is provided as supplementary material.}
	\label{fig:q_estimation_Retau180_u_PffTrue}
\end{figure}

Figure \ref{fig:q_estimation_Retau180_u_PffTrue} shows snapshot estimates for the streamwise velocity component at wall-normal planes located at $y^{+} \approx 15$ and 90 for the $Re_{\tau} \approx 180$ case.
The resemblance is remarkable, especially for the $y^{+} \approx 15$ plane.
Even at $y^{+} \approx 90$, the estimates agree quite well with the DNS snapshot.
Note that the $y^{+} \approx 90$ coordinate is equivalent to approximately $y = 0.5$, i.e., a quarter of the channel height in outer units, which is quite far from the wall.
Estimations of other state components (i.e., wall-normal and spanwise velocity components and pressure component), exhibit similar results as those shown in figures \ref{fig:q_estimation_Retau550_yplus15}, \ref{fig:q_estimation_Retau550_yplus100} and \ref{fig:q_estimation_Retau550_yplus200}.

\begin{figure}
	\centerline{\includegraphics[width=0.7\textwidth]{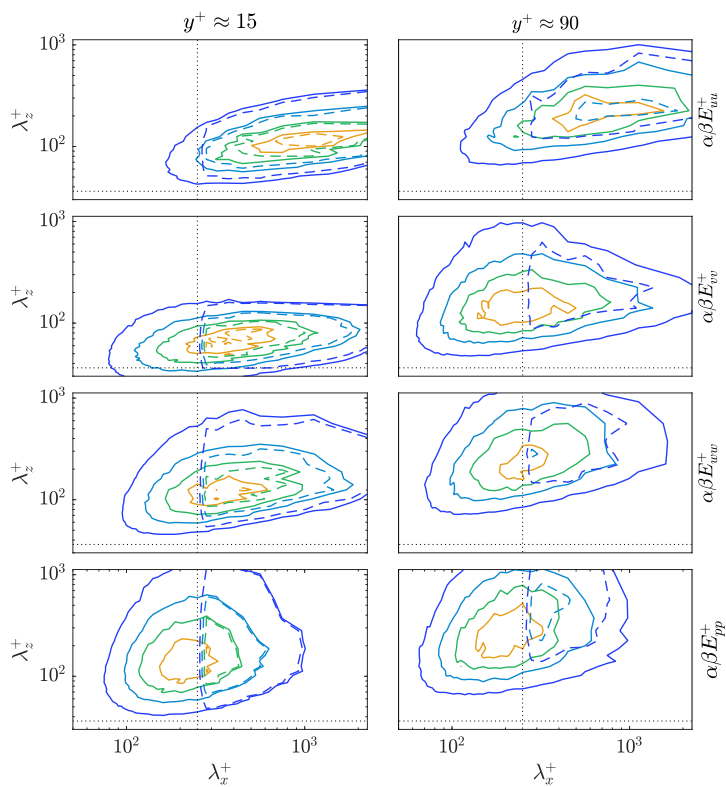}}
	\caption{Premultiplied power spectra comparison between DNS (continuous lines) and resolvent-based estimates using true forcing statistics (dashed lines) and wall measurements of shear stress and pressure for $Re_{\tau} \approx 180$. Left column: $y^{+} \approx 15$. Right column: $y^{+} \approx 90$. See comments in the caption of figure \ref{fig:pre-multiplied_energy_spectra_Retau550_PffTrue}.}
	\label{fig:pre-multiplied_energy_spectra_Retau180_PffTrue}
\end{figure}

\begin{figure}
	\centerline{\includegraphics[width=0.7\textwidth]{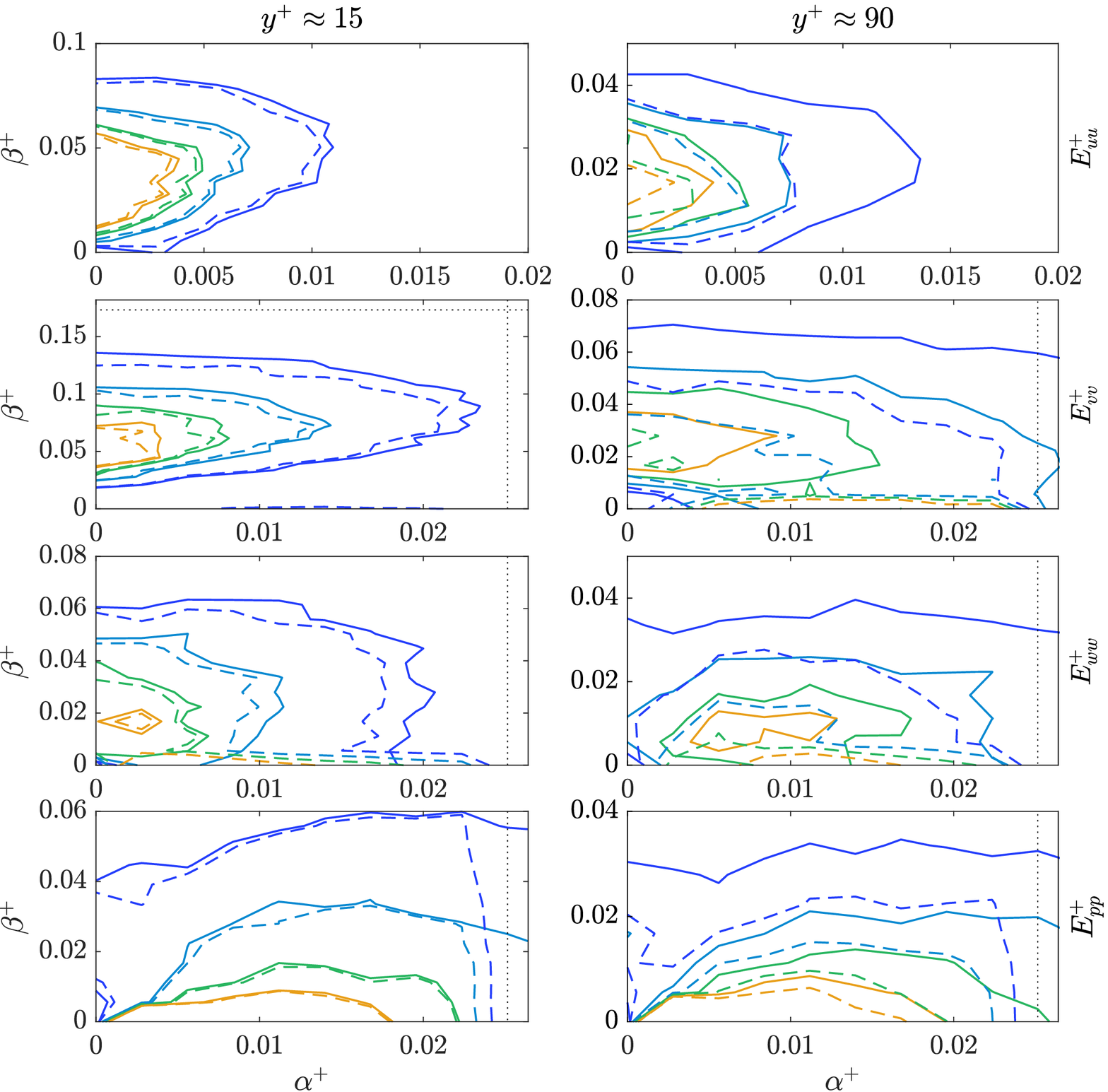}}
	\caption{Power spectra comparison between DNS (continuous lines) and resolvent-based estimates using true forcing statistics (dashed lines) and wall measurements of shear stress and pressure for $Re_{\tau} \approx 180$. See comments in the caption of figure \ref{fig:energy_spectra_Retau550_PffTrue}.}
	\label{fig:energy_spectra_Retau180_PffTrue}
\end{figure}

Power spectra of the state components with and without premultiplication are displayed in figures \ref{fig:pre-multiplied_energy_spectra_Retau180_PffTrue} and \ref{fig:energy_spectra_Retau180_PffTrue}, respectively, for wall-normal distances of $y^{+} \approx 15$ and 90.
Continuous lines denote DNS results and dashed lines the estimates using the true forcing assumption.
Good comparisons are obtained near the wall ($y^{+} \approx 15$) for all state components, and farther from the wall the results are worse, following trends seen for the other Reynolds numbers.

\begin{figure}
	\centerline{\includegraphics[width=\textwidth]{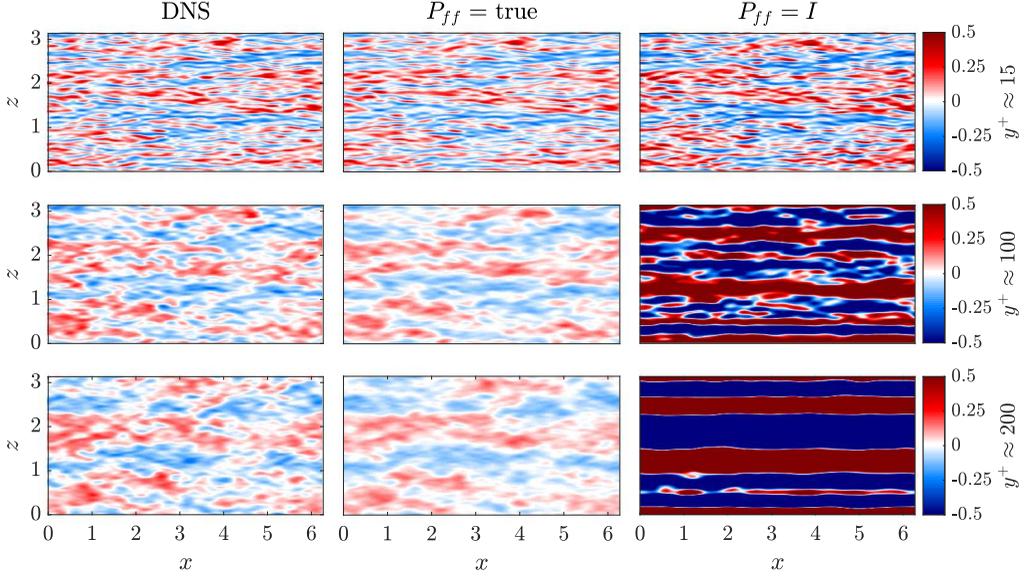}}
	\caption{Comparison between filtered DNS and resolvent-based estimates of a snapshot of the flow streamwise velocity for $Re_{\tau} \approx 1000$ and wall measurements of shear stress. Rows, from top to bottom: $y^{+} \approx 15$, 100 and 200. See comments in the caption of figure \ref{fig:q_estimation_Retau180_u_PffTrue}.}
	\label{fig:q_estimation_Retau1000_u_PffTrue}
\end{figure}

Figure \ref{fig:q_estimation_Retau1000_u_PffTrue} shows the streamwise velocity component estimates at wall-normal positions of $y^{+} \approx 15$, 100 and 200 for the $Re_{\tau} \approx 1000$ case.
DNS snapshots are also shown to enable direct comparisons, and good estimates are observed up to $y^{+} \approx 200$.
Most of the large-scale structures present in the DNS were captured by the estimations.
Reconstructions of wall-normal and spanwise velocity components are of similar quality as those obtained for the $Re_{\tau} \approx 550$ case and will not be shown here for brevity.
When using white noise forcing, we obtained estimates of similar quality to those provided by the $Re_{\tau} \approx 550$ case, i.e., the near-wall region was well estimated whereas far from the wall the snapshots were poorly estimated.

\begin{figure}
	\centerline{\includegraphics[width=\textwidth]{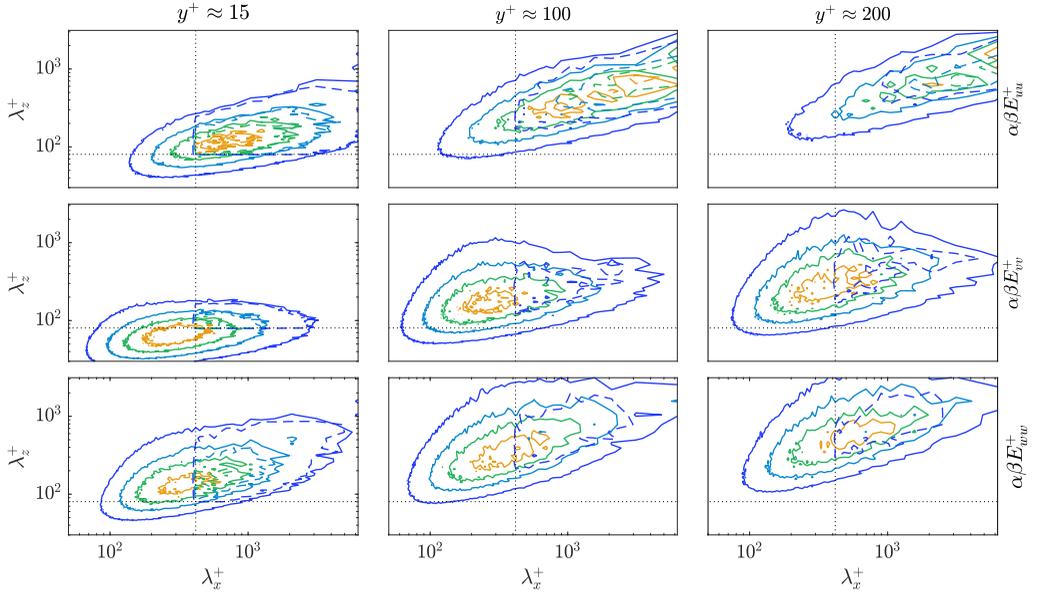}}
	\caption{Premultiplied power spectra comparison between DNS (continuous lines) and resolvent-based estimates using true forcing statistics (dashed lines), using wall measurements of shear stress for $Re_{\tau} \approx 1000$. Columns, from left to right: $y^{+} \approx 15$, 100 and 200. See comments in the caption of figure \ref{fig:pre-multiplied_energy_spectra_Retau550_PffTrue}.}
	\label{fig:pre-multiplied_energy_spectra_Retau1000_PffTrue}
\end{figure}

\begin{figure}
	\centerline{\includegraphics[width=\textwidth]{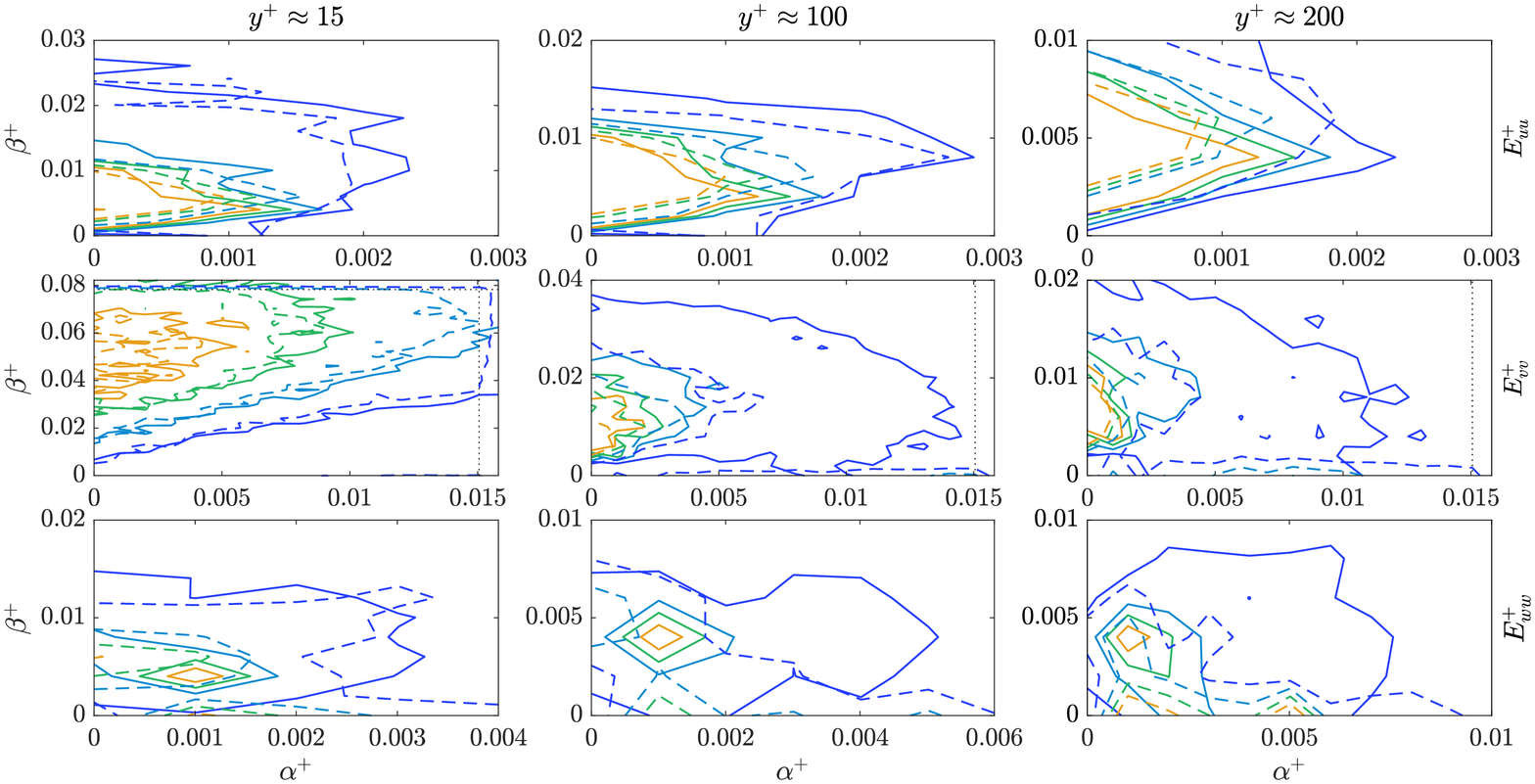}}
	\caption{Power spectra comparison between DNS (continuous lines) and resolvent-based estimates considering true forcing statistics (dashed lines) and wall measurements of shear stress for $Re_{\tau} \approx 1000$. See comments in the caption of figure \ref{fig:energy_spectra_Retau550_PffTrue}.}
	\label{fig:energy_spectra_Retau1000_PffTrue}
\end{figure}

Figures \ref{fig:pre-multiplied_energy_spectra_Retau1000_PffTrue} and \ref{fig:energy_spectra_Retau1000_PffTrue} show the power spectra of velocity fluctuations with and without premultiplication, respectively, at wall-normal positions of $y^{+} \approx 15$, 100 and 200.
The estimations (dashed lines) were performed considering true forcing statistics.
DNS (continuous lines) results are also shown in the figure to enable direct comparisons.
As in figures \ref{fig:pre-multiplied_energy_spectra_Retau550_PffTrue} and \ref{fig:energy_spectra_Retau550_PffTrue}, the estimates were normalised by the peak value of the DNS power spectra and the upper right quadrant formed by the intersection between the dotted lines indicates the wavenumbers retained in the filtered data (table \ref{tab:dns_parameters}).

In the near-wall region, $y^{+} \approx 15$, the resolvent-based estimates recovered the premultiplied power spectra with good accuracy, especially when considering the streamwise velocity component.
For the wall-normal and spanwise velocity components, the premultiplied power spectra peaks lie outside the observable region, owing to the spatial filter employed.
Nevertheless, the observable portion of the power spectra are in good agreement with DNS results.
Farther from the wall, the estimates deteriorate, as expected.
It should be emphasised that spectra without premultiplication highlight the largest scales in the flow.

\bibliographystyle{abbrvjfm}
\bibliography{references}

\begin{thebibliography}{55}
\expandafter\ifx\csname natexlab\endcsname\relax\def\natexlab#1{#1}\fi
\def\au#1{#1} \def\ed#1{#1} \def\yr#1{#1}\def\at#1{#1}\def\jt#1{\textit{#1}}
  \def\bt#1{#1}\def\bvol#1{\textbf{#1}} \def\vol#1{#1} \def\pg#1{#1}
  \def\publ#1{#1}\def\arxiv#1{#1}\def\org#1{#1}\def\st#1{\textit{#1}}

\bibitem[Abreu {\em et~al.\/}(2020{\natexlab{{\em a\/}}})Abreu, Cavalieri,
  Schlatter, Vinuesa \& Henningson]{abreu2020resolvent}
{\sc \au{Abreu, L.~I. }, \au{Cavalieri, A. V.~G. }, \au{Schlatter, P. },
  \au{Vinuesa, R. } \& \au{Henningson, D.~S. }} \yr{2020{\natexlab{{\em a\/}}}}
   \at{Resolvent modelling of near-wall coherent structures in turbulent
  channel flow}.  \jt{{International Journal of Heat and Fluid Flow}}
  \bvol{85},  \pg{108662}.

\bibitem[Abreu {\em et~al.\/}(2020{\natexlab{{\em b\/}}})Abreu, Cavalieri,
  Schlatter, Vinuesa \& Henningson]{abreu2020spod}
{\sc \au{Abreu, L.~I. }, \au{Cavalieri, A. V.~G. }, \au{Schlatter, P. },
  \au{Vinuesa, R. } \& \au{Henningson, D.~S. }} \yr{2020{\natexlab{{\em b\/}}}}
   \at{Spectral proper orthogonal decomposition and resolvent analysis of
  near-wall coherent structures in turbulent pipe flows}.  \jt{{Journal of
  Fluid Mechanics}}  \bvol{900},  \pg{A11}.

\bibitem[Abreu {\em et~al.\/}(2017)Abreu, Cavalieri \& Wolf]{abreu2017coherent}
{\sc \au{Abreu, L.~I. }, \au{Cavalieri, A. V.~G. } \& \au{Wolf, W. }} \yr{2017}
  Coherent hydrodynamic waves and trailing-edge noise.  \bt{In {\em {23rd
  AIAA/CEAS Aeroacoustics Conference}\/}},  \pg{p. 3173}.

\bibitem[Anantharamu \& Mahesh(2020)]{anantharamu2020analysis}
{\sc \au{Anantharamu, S. } \& \au{Mahesh, K. }} \yr{2020}  \at{Analysis of
  wall-pressure fluctuation sources from direct numerical simulation of
  turbulent channel flow}.  \jt{Journal of Fluid Mechanics}  \bvol{898}~(A17).

\bibitem[Bagheri {\em et~al.\/}(2009)Bagheri, Henningson, Hoepffner \&
  Schmid]{bagheri2009input}
{\sc \au{Bagheri, S. }, \au{Henningson, D.~S. }, \au{Hoepffner, J. } \&
  \au{Schmid, P.~J. }} \yr{2009}  \at{Input-output analysis and control design
  applied to a linear model of spatially developing flows}.  \jt{{Applied
  Mechanics Reviews}}  \bvol{62}~(2).

\bibitem[Beneddine {\em et~al.\/}(2016)Beneddine, Sipp, Arnault, Dandois \&
  Lesshafft]{beneddine2016conditions}
{\sc \au{Beneddine, S. }, \au{Sipp, D. }, \au{Arnault, A. }, \au{Dandois, J. }
  \& \au{Lesshafft, L. }} \yr{2016}  \at{Conditions for validity of mean flow
  stability analysis}.  \jt{{Journal of Fluid Mechanics}}  \bvol{798},
  \pg{485--504}.

\bibitem[Bewley \& Protas(2004)]{bewley2004skin}
{\sc \au{Bewley, T.~R. } \& \au{Protas, B. }} \yr{2004}  \at{Skin friction and
  pressure: the “footprints” of turbulence}.  \jt{Physica D: Nonlinear
  Phenomena}  \bvol{196}~(1-2),  \pg{28--44}.

\bibitem[Cavalieri {\em et~al.\/}(2019)Cavalieri, Jordan \&
  Lesshafft]{cavalieri2019wave}
{\sc \au{Cavalieri, A. V.~G. }, \au{Jordan, P. } \& \au{Lesshafft, L. }}
  \yr{2019}  \at{Wave-packet models for jet dynamics and sound radiation}.
  \jt{{Applied Mechanics Reviews}}  \bvol{71},  \pg{020802(27)}.

\bibitem[Chevalier {\em et~al.\/}(2006)Chevalier, H{\oe}pffner, Bewley \&
  Henningson]{chevalier2006state}
{\sc \au{Chevalier, M. }, \au{H{\oe}pffner, J. }, \au{Bewley, T.~R. } \&
  \au{Henningson, D.~S. }} \yr{2006}  \at{{State estimation in wall-bounded
  flow systems. Part 2. Turbulent flows}}.  \jt{Journal of Fluid Mechanics}
  \bvol{552}~(1),  \pg{167}.

\bibitem[Colburn {\em et~al.\/}(2011)Colburn, Cessna \&
  Bewley]{colburn2011state}
{\sc \au{Colburn, C.~H. }, \au{Cessna, J.~B. } \& \au{Bewley, T.~R. }}
  \yr{2011}  \at{{State estimation in wall-bounded flow systems. Part 3. The
  ensemble Kalman filter}}.  \jt{Journal of Fluid Mechanics}  \bvol{682},
  \pg{289--303}.

\bibitem[Del~{\'A}lamo \& Jim{\'e}nez(2003)]{delalamo2003spectra}
{\sc \au{Del~{\'A}lamo, J.~C. } \& \au{Jim{\'e}nez, J. }} \yr{2003}
  \at{Spectra of the very large anisotropic scales in turbulent channels}.
  \jt{Physics of Fluids}  \bvol{15}~(6),  \pg{L41}.

\bibitem[Del~{\'A}lamo \& Jim{\'e}nez(2006)]{del2006linear}
{\sc \au{Del~{\'A}lamo, J.~C. } \& \au{Jim{\'e}nez, J. }} \yr{2006}
  \at{{Linear energy amplification in turbulent channels}}.  \jt{Journal of
  Fluid Mechanics}  \bvol{559},  \pg{205}.

\bibitem[Del~{\'A}lamo {\em et~al.\/}(2004)Del~{\'A}lamo, Jim{\'e}nez,
  Zandonade \& Moser]{delalamo2004scaling}
{\sc \au{Del~{\'A}lamo, J.~C. }, \au{Jim{\'e}nez, J. }, \au{Zandonade, P. } \&
  \au{Moser, R.~D. }} \yr{2004}  \at{Scaling of the energy spectra of turbulent
  channels}.  \jt{Journal of Fluid Mechanics}  \bvol{500},  \pg{135}.

\bibitem[Encinar \& Jim{\'e}nez(2019)]{encinar2019logarithmic}
{\sc \au{Encinar, M.~P. } \& \au{Jim{\'e}nez, J. }} \yr{2019}
  \at{{Logarithmic-layer turbulence: A view from the wall}}.  \jt{Physical
  Review Fluids}  \bvol{4}~(11),  \pg{114603}.

\bibitem[Farrell \& Ioannou(2012)]{farrell2012dynamics}
{\sc \au{Farrell, B.~F. } \& \au{Ioannou, P.~J. }} \yr{2012}  \at{{Dynamics of
  streamwise rolls and streaks in turbulent wall-bounded shear flow}}.
  \jt{Journal of Fluid Mechanics}  \bvol{708},  \pg{149}.

\bibitem[Gibson {\em et~al.\/}(2019)Gibson, Reetz, Azimi, Ferraro, Kreilos,
  Schrobsdorff, Farano, Yesil, Sch\"{u}tz, Culpo \&
  Schneider]{gibson2019channelflow}
{\sc \au{Gibson, J.~F. }, \au{Reetz, F. }, \au{Azimi, S. }, \au{Ferraro, A. },
  \au{Kreilos, T. }, \au{Schrobsdorff, H. }, \au{Farano, M. }, \au{Yesil, A.~F.
  }, \au{Sch\"{u}tz, S.~S. }, \au{Culpo, M. } \& \au{Schneider, T.~M. }}
  \yr{2019}  \at{{ChannelFlow 2.0}}.  \jt{{Manuscript in preparation}} .

\bibitem[Guastoni {\em et~al.\/}(2020)Guastoni, G{\"u}emes, Ianiro, Discetti,
  Schlatter, Azizpour \& Vinuesa]{guastoni2020convolutional}
{\sc \au{Guastoni, L. }, \au{G{\"u}emes, A. }, \au{Ianiro, A. }, \au{Discetti,
  S. }, \au{Schlatter, P. }, \au{Azizpour, H. } \& \au{Vinuesa, R. }} \yr{2020}
   \at{Convolutional-network models to predict wall-bounded turbulence from
  wall quantities}.  \jt{arXiv:2006.12483} .

\bibitem[H{\oe}pffner {\em et~al.\/}(2005)H{\oe}pffner, Chevalier, Bewley \&
  Hennington]{hoepffner2005state}
{\sc \au{H{\oe}pffner, J. }, \au{Chevalier, M. }, \au{Bewley, T.~R. } \&
  \au{Hennington, D.~S. }} \yr{2005}  \at{{State estimation in wall-bounded
  flow systems. Part 1. Perturbed laminar flows}}.  \jt{Journal of Fluid
  Mechanics}  \bvol{534},  \pg{263--294}.

\bibitem[Hwang \& Cossu(2010)]{hwang2010linear}
{\sc \au{Hwang, Y. } \& \au{Cossu, C. }} \yr{2010}  \at{Linear non-normal
  energy amplification of harmonic and stochastic forcing in the turbulent
  channel flow}.  \jt{Journal of Fluid Mechanics}  \bvol{664},  \pg{51--73}.

\bibitem[Illingworth {\em et~al.\/}(2018)Illingworth, Monty \&
  Marusic]{illingworth2018estimating}
{\sc \au{Illingworth, S.~J. }, \au{Monty, J.~P. } \& \au{Marusic, I. }}
  \yr{2018}  \at{Estimating large-scale structures in wall turbulence using
  linear models}.  \jt{Journal of Fluid Mechanics}  \bvol{842},  \pg{146--162}.

\bibitem[Jeun {\em et~al.\/}(2016)Jeun, Nichols \&
  Jovanovi{\'c}]{jeun2016input}
{\sc \au{Jeun, J. }, \au{Nichols, J.~W. } \& \au{Jovanovi{\'c}, M.~R. }}
  \yr{2016}  \at{Input-output analysis of high-speed axisymmetric isothermal
  jet noise}.  \jt{{Physics of Fluids}}  \bvol{28}~(4),  \pg{047101}.

\bibitem[Jim{\'e}nez(2013)]{jimenez2013near}
{\sc \au{Jim{\'e}nez, J. }} \yr{2013}  \at{Near-wall turbulence}.  \jt{Physics
  of Fluids}  \bvol{25}~(10),  \pg{101302}.

\bibitem[Jung {\em et~al.\/}(2020)Jung, Martini, Cavalieri, Jordan, Lesshafft
  \& Towne]{jung2020optimal}
{\sc \au{Jung, J. }, \au{Martini, E. }, \au{Cavalieri, A. }, \au{Jordan, P. },
  \au{Lesshafft, L. } \& \au{Towne, A. }} \yr{2020}  \at{Optimal
  resolvent-based estimation for flow control}.  \jt{Bulletin of the American
  Physical Society} .

\bibitem[Lee \& Moser(2015)]{lee2015direct}
{\sc \au{Lee, M. } \& \au{Moser, R.~D. }} \yr{2015}  \at{{Direct numerical
  simulation of turbulent channel flow up to $Re_{\tau} \approx 5200$}}.
  \jt{Journal of Fluid Mechanics}  \bvol{774},  \pg{395--415}.

\bibitem[Lesshafft {\em et~al.\/}(2019)Lesshafft, Semeraro, Jaunet, Cavalieri
  \& Jordan]{lesshafft2019resolvent}
{\sc \au{Lesshafft, L. }, \au{Semeraro, O. }, \au{Jaunet, V. }, \au{Cavalieri,
  A. V.~G. } \& \au{Jordan, P. }} \yr{2019}  \at{Resolvent-based modeling of
  coherent wave packets in a turbulent jet}.  \jt{Physical Review Fluids}
  \bvol{4}~(6),  \pg{063901}.

\bibitem[Lozano-Dur{\'a}n \& Jim{\'e}nez(2014)]{lozanoduran2014effect}
{\sc \au{Lozano-Dur{\'a}n, A. } \& \au{Jim{\'e}nez, J. }} \yr{2014}
  \at{{Effect of the computational domain on direct simulations of turbulent
  channels up to Re$\tau$= 4200}}.  \jt{Physics of Fluids}  \bvol{26}~(1),
  \pg{011702}.

\bibitem[Martinelli(2009)]{martinelli2009feedback}
{\sc \au{Martinelli, F. }} \yr{2009}  \at{Feedback control of turbulent wall
  flows}. PhD thesis, Politecnico di Milano.

\bibitem[Martini(2019)]{martini2019linear}
{\sc \au{Martini, E. }} \yr{2019}  \at{Linear modelling of flows: physical
  mechanisms and tools}. PhD thesis, Instituto Tecnol{\'o}gico de
  Aeron{\'a}utica.

\bibitem[Martini {\em et~al.\/}(2020)Martini, Jordan, Cavalieri, Towne \&
  Lesshafft]{martini2020resolvent}
{\sc \au{Martini, E. }, \au{Jordan, P. }, \au{Cavalieri, A. V.~G. }, \au{Towne,
  A. } \& \au{Lesshafft, L. }} \yr{2020}  \at{Resolvent-based optimal
  estimation of transitional and turbulent flows}.  \jt{{Journal of Fluid
  Mechanics}}  \bvol{900},  \pg{A2}.

\bibitem[Marusic {\em et~al.\/}(2017)Marusic, Baars \&
  Hutchins]{marusic2017scaling}
{\sc \au{Marusic, I. }, \au{Baars, W.~J. } \& \au{Hutchins, N. }} \yr{2017}
  \at{Scaling of the streamwise turbulence intensity in the context of
  inner-outer interactions in wall turbulence}.  \jt{Physical Review Fluids}
  \bvol{2}~(10),  \pg{100502}.

\bibitem[Marusic {\em et~al.\/}(2010)Marusic, Mathis \&
  Hutchins]{marusic2010predictive}
{\sc \au{Marusic, I. }, \au{Mathis, R. } \& \au{Hutchins, N. }} \yr{2010}
  \at{Predictive model for wall-bounded turbulent flow}.  \jt{Science}
  \bvol{329}~(5988),  \pg{193--196}.

\bibitem[McKeon \& Sharma(2010)]{mckeon2010critical}
{\sc \au{McKeon, B.~J. } \& \au{Sharma, A.~S. }} \yr{2010}  \at{A
  critical-layer framework for turbulent pipe flow}.  \jt{{Journal of Fluid
  Mechanics}}  \bvol{658},  \pg{336--382}.

\bibitem[Meditch(1973)]{meditch1973survey}
{\sc \au{Meditch, J.~S. }} \yr{1973}  \at{A survey of data smoothing for linear
  and nonlinear dynamic systems}.  \jt{Automatica}  \bvol{9}~(2),
  \pg{151--162}.

\bibitem[Morra {\em et~al.\/}(2021)Morra, Nogueira, Cavalieri \&
  Henningson]{morra2021colour}
{\sc \au{Morra, P. }, \au{Nogueira, P. A.~S. }, \au{Cavalieri, A. V.~G. } \&
  \au{Henningson, D.~S. }} \yr{2021}  \at{The colour of forcing statistics in
  resolvent analyses of turbulent channel flows}.  \jt{{Journal of Fluid
  Mechanics}}  \bvol{907},  \pg{A24}.

\bibitem[Morra {\em et~al.\/}(2020)Morra, Sasaki, Hanifi, Cavalieri \&
  Henningson]{morra2020realizable}
{\sc \au{Morra, P. }, \au{Sasaki, K. }, \au{Hanifi, A. }, \au{Cavalieri, A.
  V.~G. } \& \au{Henningson, D.~S. }} \yr{2020}  \at{A realizable data-driven
  approach to delay bypass transition with control theory}.  \jt{{Journal of
  Fluid Mechanics}}  \bvol{883},  \pg{A33}.

\bibitem[Morra {\em et~al.\/}(2019)Morra, Semeraro, Henningson \&
  Cossu]{morra2019relevance}
{\sc \au{Morra, P. }, \au{Semeraro, O. }, \au{Henningson, D.~S. } \& \au{Cossu,
  C. }} \yr{2019}  \at{{On the relevance of Reynolds stresses in resolvent
  analyses of turbulent wall-bounded flows}}.  \jt{{Journal of Fluid
  Mechanics}}  \bvol{867},  \pg{969--984}.

\bibitem[Oehler {\em et~al.\/}(2018)Oehler, Garcia-Guti{\'e}rrez \&
  Illingworth]{oehler2018linear}
{\sc \au{Oehler, S. }, \au{Garcia-Guti{\'e}rrez, A. } \& \au{Illingworth, S. }}
  \yr{2018}  \at{Linear estimation of coherent structures in wall-bounded
  turbulence at re$\tau$= 2000}.  \jt{Journal of Physics: Conference Series}
  \bvol{1001}~(1),  \pg{012006}.

\bibitem[Pickering {\em et~al.\/}(2021)Pickering, Rigas, Schmidt, Sipp \&
  Colonius]{pickering2021optimal}
{\sc \au{Pickering, E. }, \au{Rigas, G. }, \au{Schmidt, O.~T. }, \au{Sipp, D. }
  \& \au{Colonius, T. }} \yr{2021}  \at{Optimal eddy viscosity for
  resolvent-based models of coherent structures in turbulent jets}.
  \jt{{Journal of Fluid Mechanics}}  \bvol{917},  \pg{A29}.

\bibitem[Pujals {\em et~al.\/}(2009)Pujals, Garc{\'\i}a-Villalba, Cossu \&
  Depardon]{pujals2009note}
{\sc \au{Pujals, G. }, \au{Garc{\'\i}a-Villalba, M. }, \au{Cossu, C. } \&
  \au{Depardon, S. }} \yr{2009}  \at{A note on optimal transient growth in
  turbulent channel flows}.  \jt{Physics of Fluids}  \bvol{21}~(1),
  \pg{015109}.

\bibitem[Reynolds \& Tiederman(1967)]{reynolds1967stability}
{\sc \au{Reynolds, W.~C. } \& \au{Tiederman, W.~G. }} \yr{1967}  \at{{Stability
  of turbulent channel flow, with application to Malkus's theory}}.
  \jt{Journal of Fluid Mechanics}  \bvol{27}~(2),  \pg{253--272}.

\bibitem[Sasaki {\em et~al.\/}(2017)Sasaki, Piantanida, Cavalieri \&
  Jordan]{sasaki2017real}
{\sc \au{Sasaki, K. }, \au{Piantanida, S. }, \au{Cavalieri, A. V.~G. } \&
  \au{Jordan, P. }} \yr{2017}  \at{Real-time modelling of wavepackets in
  turbulent jets}.  \jt{Journal of Fluid Mechanics}  \bvol{821},
  \pg{458--481}.

\bibitem[Sasaki {\em et~al.\/}(2019)Sasaki, Vinuesa, Cavalieri, Schlatter \&
  Henningson]{sasaki2019transfer}
{\sc \au{Sasaki, K. }, \au{Vinuesa, R. }, \au{Cavalieri, A. V.~G. },
  \au{Schlatter, P. } \& \au{Henningson, D.~S. }} \yr{2019}  \at{Transfer
  functions for flow predictions in wall-bounded turbulence}.  \jt{Journal of
  Fluid Mechanics}  \bvol{864},  \pg{708--745}.

\bibitem[Schmidt {\em et~al.\/}(2018)Schmidt, Towne, Rigas, Colonius \&
  Br{\'e}s]{schmidt2018spectral}
{\sc \au{Schmidt, O.~T. }, \au{Towne, A. }, \au{Rigas, G. }, \au{Colonius, T. }
  \& \au{Br{\'e}s, G.~A. }} \yr{2018}  \at{Spectral analysis of jet
  turbulence}.  \jt{Journal of Fluid Mechanics}  \bvol{855},  \pg{953--982}.

\bibitem[Sharma \& McKeon(2013)]{sharma2013coherent}
{\sc \au{Sharma, A.~S. } \& \au{McKeon, B.~J. }} \yr{2013}  \at{On coherent
  structure in wall turbulence}.  \jt{{Journal of Fluid Mechanics}}
  \bvol{728},  \pg{196--238}.

\bibitem[Smits {\em et~al.\/}(2011)Smits, McKeon \& Marusic]{smits2011high}
{\sc \au{Smits, A.~J. }, \au{McKeon, B.~J. } \& \au{Marusic, I. }} \yr{2011}
  \at{{High-Reynolds number wall turbulence}}.  \jt{Annual Review of Fluid
  Mechanics}  \bvol{43},  \pg{353--375}.

\bibitem[Symon {\em et~al.\/}(2021)Symon, Illingworth \&
  Marusic]{symon2021energy}
{\sc \au{Symon, S. }, \au{Illingworth, S.~J. } \& \au{Marusic, I. }} \yr{2021}
  \at{Energy transfer in turbulent channel flows and implications for resolvent
  modelling}.  \jt{{Journal of Fluid Mechanics}}  \bvol{911},  \pg{A3}.

\bibitem[Taira {\em et~al.\/}(2017)Taira, Brunton, Dawson, Rowley, Colonius,
  McKeon, Schmidt, Gordeyev, Theofilis \& Ukeiley]{taira2017modal}
{\sc \au{Taira, K. }, \au{Brunton, S.~L. }, \au{Dawson, S. T.~M. }, \au{Rowley,
  C.~W. }, \au{Colonius, T. }, \au{McKeon, B.~J. }, \au{Schmidt, O.~T. },
  \au{Gordeyev, S. }, \au{Theofilis, V. } \& \au{Ukeiley, L.~S. }} \yr{2017}
  \at{{Modal analysis of fluid flows: An overview}}.  \jt{{AIAA Journal}}
  \bvol{55}~(12),  \pg{4013--4041}.

\bibitem[Thomas {\em et~al.\/}(2015)Thomas, Farrell, Ioannou \&
  Gayme]{thomas2015minimal}
{\sc \au{Thomas, V.~L. }, \au{Farrell, B.~F. }, \au{Ioannou, P.~J. } \&
  \au{Gayme, D.~F. }} \yr{2015}  \at{{A minimal model of self-sustaining
  turbulence}}.  \jt{Physics of Fluids}  \bvol{27}~(10),  \pg{105104}.

\bibitem[Tinney {\em et~al.\/}(2006)Tinney, Coiffet, Delville, Hall, Jordan \&
  Glauser]{tinney2006spectral}
{\sc \au{Tinney, C.~E. }, \au{Coiffet, F. }, \au{Delville, J. }, \au{Hall,
  A.~M. }, \au{Jordan, P. } \& \au{Glauser, M.~N. }} \yr{2006}  \at{On spectral
  linear stochastic estimation}.  \jt{{Experiments in Fluids}}  \bvol{41}~(5),
  \pg{763--775}.

\bibitem[Tissot {\em et~al.\/}(2017)Tissot, Zhang, Laj{\'u}s, Cavalieri \&
  Jordan]{tissot2017sensitivity}
{\sc \au{Tissot, G. }, \au{Zhang, M. }, \au{Laj{\'u}s, F.~C. }, \au{Cavalieri,
  A. V.~G. } \& \au{Jordan, P. }} \yr{2017}  \at{Sensitivity of wavepackets in
  jets to nonlinear effects: the role of the critical layer}.  \jt{{Journal of
  Fluid Mechanics}}  \bvol{811},  \pg{95--137}.

\bibitem[Tol {\em et~al.\/}(2019)Tol, de \& Kotsonis]{tol2019experimental}
{\sc \au{Tol, H.~J. }, \au{de, Visser, C.~C. } \& \au{Kotsonis, M. }} \yr{2019}
   \at{{Experimental Model-Based Estimation and Control of Natural
  Tollmien--Schlichting Waves}}.  \jt{AIAA Journal}  \bvol{57}~(6),
  \pg{2344--2355}.

\bibitem[Towne {\em et~al.\/}(2017)Towne, Br{\'e}s \&
  Lele]{towne2017statistical}
{\sc \au{Towne, A. }, \au{Br{\'e}s, G.~A. } \& \au{Lele, S.~K. }} \yr{2017} A
  statistical jet-noise model based on the resolvent framework.  \bt{In {\em
  {23rd AIAA/CEAS Aeroacoustics Conference}\/}},  \pg{p. 3706}.

\bibitem[Towne {\em et~al.\/}(2020)Towne, Lozando-Dur{\'a}n \&
  Yang]{towne2020resolvent}
{\sc \au{Towne, A. }, \au{Lozando-Dur{\'a}n, A. } \& \au{Yang, X. }} \yr{2020}
  \at{Resolvent-based estimation of space–time flow statistics}.
  \jt{{Journal of Fluid Mechanics}}  \bvol{883},  \pg{A17}.

\bibitem[Towne {\em et~al.\/}(2018)Towne, Schmidt \&
  Colonius]{towne2018spectral}
{\sc \au{Towne, A. }, \au{Schmidt, O.~T. } \& \au{Colonius, T. }} \yr{2018}
  \at{Spectral proper orthogonal decomposition and its relationship to dynamic
  mode decomposition and resolvent analysis}.  \jt{{Journal of Fluid
  Mechanics}}  \bvol{847},  \pg{821--867}.

\bibitem[Welch(1967)]{welch1967fft}
{\sc \au{Welch, P.~D. }} \yr{1967}  \at{{The use of Fast Fourier Transform for
  the estimation of power spectra: A method based on time averaging over short,
  modified periodograms}}.  \jt{IEEE Transactions on Audio and
  Electroacoustics}  \bvol{15}~(2),  \pg{70--73}.

\end{thebibliography}

\end{document}